\documentclass[11pt,a4paper]{article}
\pdfoutput=1

\usepackage{jheppub}
\usepackage[rgb]{xcolor}
\usepackage{color}
\usepackage{amsmath,amsfonts,mathrsfs,graphicx,bbm,dsfont,booktabs,mathtools,braket,lmodern,tikz}

\DeclareMathAlphabet{\mathpzc}{OT1}{pzc}{m}{it}

\def\L{{\mathbb L}}

\newcommand{\bea}{\begin{eqnarray}}
\newcommand{\eea}{\end{eqnarray}}
\def\be{\begin{equation}}
\def\ee{\end{equation}}
\newcommand{\bei}{\begin{itemize}}
\newcommand{\eei}{\end{itemize}}
\newcommand{\bee}{\begin{enumerate}}
\newcommand{\eee}{\end{enumerate}}

\def\a {\alpha}
\def\b {\beta}

\def\eps{\epsilon}

\def\ov{\over}

\def\pa {\partial}

\def\L{\mathscr L}

\newcommand{\alg}[1]{\mathfrak{#1}}
\newcommand{\su}{\alg{su}}

\newcommand{\psu}{\alg{psu}}

\def\ads{{\rm AdS}_5\times {\rm S}^5}

\def\ads{{\rm AdS}_5\times {\rm S}^5}

\def\dn{{\rm dn}}

\def\am{{\rm am}}
\def\am0{{\rm am}_0}

\DeclareMathOperator{\arcsinh}{arcsinh}

\expandafter\def\expandafter\bfseries\expandafter{\bfseries\ifmmode\else\boldmath\fi}
\expandafter\def\expandafter\mdseries\expandafter{\mdseries\ifmmode\else\unboldmath\fi}
\expandafter\def\expandafter\normalfont\expandafter{\normalfont\ifmmode\else\unboldmath\fi}

\renewcommand{\r}{{\vartheta}}
\def\rd{\tilde{\r}}
\def\agauge{{a_g}}
\def\s {\mathbf{s}}

\definecolor{grey}{rgb}{0.4,0.4,0.5}
\definecolor{darkgreen}{rgb}{0,0.5,0}
\definecolor{darkred}{rgb}{0.6,0.0,0}
\definecolor{lightbrown}{rgb}{1,0.9,0.8}
\definecolor{brown}{rgb}{0.6,0.3,0.3}
\definecolor{darkblue}{rgb}{0,0,0.8}
\definecolor{darkmagenta}{rgb}{0.5,0,0.5}

\title{On the exact spectrum and mirror duality of the $\left(\ads\right)_\eta$ superstring}

\author[a,1]{Gleb Arutyunov}
\note{Correspondent fellow at Steklov Mathematical Institute, Moscow.}
\author[b]{Marius de Leeuw}
\author[c]{and Stijn J. van Tongeren}

\affiliation[a]{Institute for Theoretical Physics and Spinoza Institute, Utrecht University, Leuvenlaan 4, 3584 CE Utrecht, The Netherlands}
\affiliation[b]{ETH Z\"urich, Institut f\"ur Theoretische Physik, Wolfgang-Pauli-Str. 27, CH-8093 Zurich, Switzerland}
 \affiliation[c]{Institut f\"ur Mathematik und Institut f\"ur Physik, Humboldt-Universit\"at zu Berlin, IRIS Geb\"aude, Zum Grossen Windkanal 6, 12489 Berlin}

\emailAdd{g.e.arutyunov@uu.nl}
\emailAdd{deleeuwm@phys.ethz.ch}
\emailAdd{svantongeren@physik.hu-berlin.de}

\abstract{We discuss the spectrum of a string propagating on $\eta$-deformed $\ads$ by treating its world-sheet theory as an integrable quantum field theory. The exact S-matrix of this field theory is given by a $q$-deformation of the $\ads$ world-sheet S-matrix with real deformation parameter. By considering mirror (double Wick-rotated) versions of these world-sheet theories we give the Thermodynamic Bethe Ansatz description of their exact finite size spectra. Interestingly, this class of models maps onto itself under the mirror transformation. At the level of the string this appears to say that the light-cone world-sheet theories of strings on particular pairs of backgrounds are related by a double Wick rotation, a feature we call `mirror duality'. We provide a partial check of these statements at the level of the sigma model by considering reduced actions and their corresponding (mirror) giant magnon solutions.}

\begin{document}

\begin{flushright}\small{HU-EP-14/13\\HU-MATH-14/06\\ITP-UU-14/09\\SPIN-14/10}\end{flushright}

\maketitle

\section{Introduction}

Determining the exact spectrum of a free string propagating on a generic given background is a highly non-trivial problem.
In particular, already the maximally symmetric  $\ads$  space emerging in the canonical example of the AdS/CFT correspondence \cite{M}
results in a very complicated world-sheet theory \cite{Arutyunov:2009ga}. Fortunately however, in the case of $\ads$ it is possible to treat the model as a quantum integrable field theory, presenting an opportunity to study it using well established techniques (see \cite{Arutyunov:2009ga,Beisert:2010jr} for reviews).

\smallskip

Recently there was an interesting proposal on how to deform the string sigma model on $\ads$  in a way which maintains its classical
integrability \cite{Delduc:2013qra}.\footnote{Some earlier and related work on sigma model deformations can be found in \cite{Cherednik:1981df,Klimcik:2008eq,Klimcik:2002zj,Delduc:2013fga,Kawaguchi:2012ve,Kawaguchi:2012gp,Sfetsos:2013wia,Hollowood:2014fha}.}  The deformation is controlled by a real deformation parameter $\eta$ which is why we refer to the
corresponding model as the `$\eta$-deformed model' or as a `string on $\left(\ads\right)_\eta$'.\footnote{While the presence of $\kappa$-symmetry on the world-sheet is a quite nontrivial and stringy statement, it has thus far not been concretely established that the resulting deformed target spaces are proper string backgrounds. We will nonetheless continue to refer to these models as strings.} Our main interest lies in the fact that the deformation breaks supersymmetry as well as all non-abelian isometries, while still leaving us with an exciting possibility to solve the model exactly.

\smallskip

Another interesting aspect of the $\eta$-deformed model is related to its hidden symmetries. It is well known that in the light-cone gauge the symmetry algebra of the $\ads$ superstring
constitutes two copies of the centrally extended superalgebra $\psu(2|2)$. This superalgebra admits a natural deformation in the sense of quantum groups, denoted $\psu_q(2|2)_{c.e.}$,
where $q$ is the deformation parameter. This deformed algebra can be used to determine the $\psu_q(2|2)_{c.e.}^{\oplus2}$-invariant S-matrix \cite{BK} which in turn can be viewed
as a q-deformation of the light-cone world-sheet S-matrix of the $\ads$ superstring. A link between this S-matrix approach based on symmetries and the physical $\eta$-deformed model has been recently established in \cite{Arutyunov:2013ega}. There it was shown that in the large tension limit  ($g\to\infty$) the tree-level bosonic S-matrix arising from the $\eta$-deformed model matches perfectly with the q-deformed S-matrix provided the deformation parameter is identified as
\begin{equation*}
q = e^{-\nu/g} \, , \, \, \, \,\mbox{where} \,\,\, \nu = \frac{2\eta}{1+\eta^2}\, .
\end{equation*}
This observation supports the hypothesis that in the full quantum theory the symmetry algebra of the $\eta$-deformed model is given by two copies of $\psu_q(2|2)$ upon proper identification of $q$ in terms of $\eta$ and the string tension. Armed with this hypothesis, in the present paper we apply the Thermodynamic Bethe Ansatz (TBA) approach \cite{Zamolodchikov90} to this $\eta$-deformed model and obtain the equations governing its spectrum similarly to how this was done in the undeformed case \cite{Arutyunov:2009zu,Arutyunov:2009ur,Bombardelli:2009ns,Gromov:2009bc} and the case where the deformation parameter $q$ lies on a circle \cite{qTBAI,qTBAII,StijnsThesis}.

\smallskip

Without knowledge on how the deformation parameter $q$ depends on $\eta$ and $g$ in full quantum theory, we simply assume that $q=e^a$ and treat $a$ as a generic (real) parameter. We then introduce a second independent parameter $\r$ ($\in[0,\pi]$) as a particularly convenient function of the (renormalized) coupling constant and $q$. As such we will be treating a two-parameter family of models.

\smallskip

To derive the ground state TBA equations, we are naturally led to study mirror theories which are obtained from the original ones by interchanging the role of space and time, \emph{i.e.} by doing a double Wick rotation \cite{Zamolodchikov90}. In contrast to relativistic integrable models the emerging mirror models are stunningly different from the original string models (the dispersion relation, scattering and bound states, the definition of the physical region, etc., all differ considerably) \cite{Arutyunov:2007tc}.

\smallskip

The presence of the deformation parameter introduces a new twist in the relationship between the original models and their mirror versions. We find that if we treat our (deformed) string models as
a continuous family parameterized by $\r$, then the mirror model corresponding to a string model with parameter ${\r_0}$ is actually equivalent to a string model
with parameter $\pi-{\r_0}$. In other words, the strings in this family are equal to the mirror versions of other strings from the same family. We refer to this phenomenon as `mirror duality'. In this paper we provide some evidence for this mirror duality by studying the (exact) dispersion relation and scattering theory, as well as the semi-classical world-sheet theory. In the latter case, as a byproduct we derive both the string and the mirror giant magnon solutions, confirming the dispersion relations of the corresponding theories.

\smallskip

Coming to the TBA equations, apart from the domains of the Y-functions we find that there is no qualitative change with respect to the undeformed case. There is an infinite number of Y-functions, and all of them except $Y_{+}$ and $Y_-$, are supported on a cylinder (the interval $(-\pi,\pi]$), while $Y_{+}$ and $Y_-$ are supported on the interval $(-\r,\r)$. This distinguishes deformations with $q$ real from those with $q$ being a phase, as in the latter case the number of Y-functions can be finite \cite{qTBAI,qTBAII,StijnsThesis}.

\smallskip

The paper is organized as follows. In section 2 we summarize the relevant physical properties of the q-deformed S-matrix. In section 3 we discuss our parametrization of
the model. Section 4 is devoted to demonstrating mirror duality of the scattering theory. In section 5 we discuss mirror duality of the sigma model in the semi-classical regime and find giant magnon solutions. In section 6 we propose the TBA equations for the deformed model. Section 7 contains our conclusions and interesting avenues for further investigation. Relevant technical details are gathered in four appendices.

\section{The world-sheet S-matrix}
\label{sec:theSmatrix}
Let us begin by introducing and summarizing the relevant properties of the exact world-sheet S-matrix of the deformed theory.

As was shown in \cite{Arutyunov:2013ega}, the bosonic semi-classical two-body world-sheet S-matrix of the $\eta$-deformed sigma model coincides with the semi-classical limit of the bosonic part of the $\mathfrak{psu}_q(2|2)_{c.e.}^{\oplus2}$ invariant S-matrix constructed as\footnote{We closely follow the approach and notation of the undeformed string reviewed in \cite{Arutyunov:2009ga}.}
\begin{equation}
\label{eq:Smatrixdef}
{\mathbf S}(p_1,p_2)=S_{\mathfrak{su}(2)} S\check{\otimes} S\, ,
\end{equation}
where $S$ is the $\mathfrak{psu}_q(2|2)_{c.e.}$ invariant S-matrix \cite{BK} explicitly given in appendix \ref{app:qdefmatrixSmatrix}, and
\begin{equation}
\label{eq:Ssu2def}
S_{\mathfrak{su}(2)}(p_1,p_2)=\frac{1}{\sigma^2(p_1,p_2)}\frac{x_1^+ + \xi}{x_1^- + \xi}\frac{x_2^- + \xi}{x_2^+ + \xi}\cdot
\frac{x_1^- - x_2^+}{x_1^+ - x_2^-}\frac{1-\frac{1}{x_1^-x_2^+}}{1-\frac{1}{x_1^+x_2^-}}\, .
\end{equation}
The dressing phase $\sigma$ is a solution of the crossing equations discussed in appendix \ref{app:dressingphase}, making the full S-matrix crossing symmetric. To understand the meaning of the parameters $x^\pm$ (with a subscript to indicate the particle) and $\xi$ we recall that in this picture the world-sheet excitations transform under two copies of the fundamental short representation of $\mathfrak{psu}_q(2|2)_{c.e.}$. Now the fundamental representation of $\mathfrak{psu}_q(2|2)_{c.e.}$ can be conveniently parametrized in terms of the parameters $x^+$ and $x^-$, while the shortening condition is met provided these parameters satisfy the constraint
\begin{equation}
\label{eq:xpmconstraint}
\frac{1}{q}\left(x^+ + \frac{1}{x^+} + \xi + \frac{1}{\xi}\right) =
q\left(x^- + \frac{1}{x^-} + \xi + \frac{1}{\xi}\right)\,,
\end{equation}
where $\xi$ and $q$ can be viewed as free parameters labeling the centrally extended quantum deformed algebra. We will relate these parameters to the string tension and deformation parameter of the string background shortly.

We will take the central elements $U$ and $V$ of $\mathfrak{psu}_q(2|2)_{c.e.}$ in the fundamental representation to be related to the energy $E$ and momentum $p$ of the world-sheet excitations in the same fashion as in the phase-deformed case \cite{qTBAI}, namely
\begin{equation}
\label{eq:Eandpdef}
V^2 = q \frac{x^+}{x^-} \frac{x^- +\xi}{x^+ + \xi} \equiv q^E  \, , \hspace{20pt}
U^2 = \frac{1}{q} \frac{x^+ + \xi}{x^- + \xi}  \equiv e^{ip}\,.
\end{equation}
Also, there is a natural `coupling constant' $h$ in the algebra, related to $\xi$ and $q$ as
\begin{equation}
\xi(h,q) = -\frac{i}{2} \frac{h (q - q^{-1})}{\sqrt{1- \frac{h^2}{4} (q - q^{-1})^2}}\, .
\end{equation}
In terms of $h$ the shortening condition reads
\begin{equation}
\label{eq:shortening}
\Big(\frac{V-V^{-1}}{q-q^{-1}}\Big)^2-\frac{h^2}{4}(1-U^2V^2)(V^{-2}-U^{-2})=1\, ,
\end{equation}
which is equivalent to \eqref{eq:xpmconstraint} above via eqs. \eqref{eq:Eandpdef} and $h(\xi,q)$.

\subsection*{Semi-classics}

As indicated in the introduction, to match the world-sheet S-matrix of the deformed sigma model in the semi-classical regime, the deformation parameter $q$ needs to be real and related to the deformation parameter $\eta$ and the `effective string tension' $g$ as
\begin{equation}
\label{eq:q(eta,g)}
q = e^{-\nu/g} \, , \, \, \, \,\mbox{where} \,\,\, \nu = \frac{2\eta}{1+\eta^2}\, .
\end{equation}
This effective string tension $g$ is introduced in \cite{Arutyunov:2013ega} as a conventional normalization of the deformed sigma model action such that it agrees with the `algebraic' coupling constant $h$ introduced above,
\begin{equation}
\label{eq:gish}
g = h\, ,
\end{equation}
at least in the semi-classical regime. We should note that the identification \eqref{eq:q(eta,g)} of \cite{Arutyunov:2013ega} strictly speaking holds for all $\nu \neq 1$, as the limit $\nu \rightarrow 1$ is not directly well defined. We will come back to this point. The semi-classical dispersion relation is perfectly reproduced in the limit $g\rightarrow \infty$ via the identification \eqref{eq:Eandpdef}, provided we rescale $p \rightarrow p/g$ as usual. We will provide a further check of our exact dispersion relation at the end of section \ref{sec:sigmamodelduality} when we consider (mirror) giant magnons. Note that semi-classically
\begin{equation}
\label{eq:xisemiclassics}
\xi = i \frac{\nu}{\sqrt{1-\nu^2}}\, .
\end{equation}

\subsection*{Beyond the semi-classical regime}

We will assume that the natural identification of the exact S-matrix as well as the energy and momentum in terms of the central charges in eqs. \eqref{eq:Eandpdef} holds beyond the semi-classical regime, though we should keep  in mind that the precise identification of $q$ as a function of $g$ and $\eta$ outside the semi-classical limit is currently unknown, and not unrelatedly, that the relation between $g$ and $h$ might undergo nontrivial renormalization.\footnote{This possible nontrivial dependence of $h$ on $g$ is similar to the situation for the $\mathrm{AdS}_4 \times \mathbb{CP}^3$ sigma model \cite{Nishioka:2008gz,Gaiotto:2008cg}. It is important to note however that here the effect is due to the deformation parameter $\eta$ and must disappear in the limit $\eta \rightarrow 0$.} One important fact to note in this regard is that while $q$ being real is necessary for unitarity of the q-deformed S-matrix, it is not sufficient. In fact, if we parametrize
\begin{equation}
q = e^{-a}\, ,
\end{equation}
the S-matrix is unitary for
\begin{equation}
\label{eq:unitarityregime}
0\leq h^2 \sinh^2 a \leq1
\end{equation}
but loses its unitarity when we cross this bound. In other words, the real-$q$--deformed S-matrix is unitary for imaginary $\xi$, but not when $\xi$ is real. Interestingly, the semi-classical identification \eqref{eq:xisemiclassics} precisely covers all (positive) imaginary $\xi$, as $\nu$ is allowed to run from zero to one, but note that $\nu=1$ is a subtle point since the unitarity bound is violated for any large but finite $g$, and the bound is approached from the wrong side in the strict limit $g\rightarrow \infty$. Working under the assumption that our $q$-deformed S-matrix really represents the exact S-matrix of the deformed sigma model, it is also clear that the semi-classical relation \eqref{eq:q(eta,g)} as it stands with $g=h$ can not hold in general, as it would result in a non-unitarity S-matrix for small $g$. We presume that the precise interpolating forms of $q(\eta,g)$ and $ h(\eta,g)$ are such that unitarity is preserved, but at the moment do not have further constraints to provide. It is for example entirely possible that the appealingly simple semi-classical relation \eqref{eq:q(eta,g)} holds for general $\eta$ and $g$, but that the functional form of $h(\eta,g)$ is such that unitarity is nonetheless preserved, or vice versa that the identification $h=g$ holds as for the undeformed model, but that the semi-classical relation \eqref{eq:q(eta,g)} gets corrected. For instance, we might imagine that
\begin{equation}
a = \arcsinh \frac{\nu}{h}\, , \,\,\,\,\, \mbox{and} \,\,\,\,\, h=g\, ,
\end{equation}
which is nothing but the direct extension of the semi-classical identification of $\xi$ and $h$ in terms of $\nu$ and $g$ respectively. Without further data however, we should allow for the most generic situation and hence we will only assume that $q$ and $h$ depend on $\eta$ and $g$ in a definite and unitarity-compatible fashion. From this point on we will therefore work in terms of the `algebraic' quantities $a$ $(\log q)$ and $h$ (or $\xi$), which are closer to the relevant parameters parametrizing our problem anyway. Since semi-classically $a$ is naturally positive ($0<q<1$) we will focus on this regime in the parameter space. Some definitions are more naturally inverted for negative $a$ ($q>1$) as we will indicate where relevant. It is similarly natural to focus on positive imaginary $\xi$, though our definitions are not sensitive to this and some concepts are more elegantly expressed by allowing negative imaginary $\xi$. At the end of the next section we will see that this region covers all unique S-matrices.

\section{Parametrization}
\label{sec:parametrization}

Fundamental short representations of $\mathfrak{psu}_q(2|2)_{c.e.}$ can be identified with points on a torus \cite{qTBAI}, uniformizing the shortening condition on the central charges. Rather than working with this torus however, we can also map the most general solution of the shortening condition to a cylinder with cuts. As the unitarity constraint is equivalent to imaginarity of $\xi$, it is convenient to change variables from the unitarity-constrained $h$ and $a$ to $\r$ and $a$, where
\begin{equation}
\xi = i \tan \frac{\r}{2}\, ,
\end{equation}
meaning we parametrized
\begin{equation}
\label{eq:thetadef}
h \sinh a = \sin \frac{\r}{2}\, .
\end{equation}
We then introduce a rapidity $u$ via
\begin{align}
\label{eq:rapiditydef}
e^{i u} := - \frac{x + \frac{1}{x} + \xi +\frac{1}{\xi}}{\xi -\frac{1}{\xi}}\, ,
\end{align}
which takes values on a cylinder ($\mbox{Re}(u) \in (-\pi,\pi]$). Up to inversion this gives us two canonical $x$-functions on the cylinder, the `string' $x$-function
\begin{equation}
x_s(u)=-i \csc \r \left(e^{iu}-\cos \r - (1-e^{iu}) \sqrt{\frac{\cos u-\cos \r}{\cos u - 1}} \right) \, .
\label{eq:xs}
\end{equation}
and the `mirror' $x$-function\footnote{For $a<0$ it is more natural to denote the inverse of this function by $x_m$.}
\begin{equation}
x_m(u)=-i \csc \r \left(e^{iu}-\cos \r + (1+e^{iu}) \sqrt{\frac{\cos u-\cos \r}{\cos u + 1}} \right)\, .
\label{eq:xm}
\end{equation}
Parametrized this way it readily follows from eqn. \eqref{eq:xpmconstraint} that
\begin{equation}
x^\pm(u) = x(u\pm ia)\, .
\end{equation}
These functions have branch points at $\pm \r$, with the branch cut of $x_s$ running between these points through the origin and that of $x_m$ running through $\pi$. These functions are each others' analytic continuation through their respective cuts, and in particular they are \emph{equal} on the \emph{lower} half of the complex plane and inverse on the upper half. As in the undeformed case we only really need one type of $x$-function to describe the general solution of the constraint \eqref{eq:xpmconstraint}, taking the pair $\{x^+,x^-\}$ to be given by $\{x_{s/m}^+,x_{s/m}^-\}$, $\{x_{s/m}^+,1/x_{s/m}^-\}$, $\{1/x_{s/m}^+,x_{s/m}^-\}$, and $\{1/x_{s/m}^+,1/x_{s/m}^-\}$ respectively. Still both functions are relevant due to their differing cut structure and conjugation properties
\begin{equation}
\label{eq:xsmconjugation}
[x_s(u)]^*  =  \frac{x_s(u^*) + \xi}{x_s(u^*) \xi +1} \,, \hspace{20pt}
[x_m(u)]^*  =  \frac{x_m(u^*) \xi +1}{x_m(u^*) + \xi} \,.
\end{equation}
Let us also note that changing the sign of $\r (\xi)$ simply changes the sign of the $x$-functions, and changing the sign of $a$ trivially interchanges the values of $x^+$ and $x^-$. By rescaling $u \rightarrow a g u$, identifying $\r(a,g)$ via eqs. \eqref{eq:thetadef} and \eqref{eq:gish}, and taking the limit $a\rightarrow0^+$ these functions and their domains readily become their undeformed counterparts with shifts implemented by $\pm i/g$.\footnote{The limit $a\rightarrow0^-$ of $x_m$ gives the inverse of the conventional undeformed mirror function, for $x_s$ there is no such distinction.}  Explicit expressions for $E$ and $p$ and defined by (\ref{eq:Eandpdef}) in terms of $u$ can be found in appendix \ref{app:qdefmatrixSmatrix}.

\subsection*{Properties of the exact S-matrix}

In addition to being (physically) unitary, Hermitian analytic, and satisfying the Yang-Baxter equation, on the $u$-cylinder it is easy to see that our $q$-deformed S-matrix is invariant under a change of sign of $\r(\xi)$ at fixed $a(q)$\footnote{These statements should all be taken in the sense of the unitary equivalence discussed in appendix \ref{app:qdefmatrixSmatrix}.}
\begin{equation}
\mathbf{S}(u,v;\r,a)\simeq \mathbf{S}(u,v;-\r,a)\,,
\end{equation}
which follows from invariance of the S-matrix under a sign flip on the $x$-functions. In terms of momentum this reads
\begin{equation}
\label{eq:Sxiinvariance}
\mathbf{S}(p_1,p_2;\r,a) \simeq \mathbf{S}(p_1,p_2;-\r,a) \, .
\end{equation}
Next, noting that under inversion of $q$ at fixed $\r$ we \emph{effectively} interchange $x^+$ and $x^-$, it is not too hard to convince ourselves that
\begin{equation}
\mathbf{S}(u,v;\r,a) \simeq \mathbf{S}^{-1}(u,v;\r,-a) \, ,
\end{equation}
anywhere on the string and mirror $u$-cylinders.\footnote{Note that this transformation involves changing the sign of $a$, and hence the transformation required for the mirror transformation in the S-matrix on the right hand side. As the entire S-matrix is invariant under $x \rightarrow 1/x$ however, this subtlety is of little consequence at this particular point. It is important when identifying the mirror momentum (see section \ref{sec:mirrorduality}) however, and is the reason for the sign change in the equation just below also for the mirror theory.}
This property might look a little strange in the limit $q \rightarrow 1$ (due to our normalization of rapidities), however noting that the momentum switches sign under an inversion of $q$ we can also represent this relation as
\begin{equation}
\label{eq:Spseudoparityfixedxi}
\mathbf{S}(p_1,p_2;\r,a) \simeq \mathbf{S}^{-1}(-p_1,-p_2;\r,-a)\, .
\end{equation}
Put together with eqn. \eqref{eq:Sxiinvariance} this tells us that the S-matrix also inverts under inversion of $q$ at fixed $h$
\begin{equation}
\label{eq:Spseudoparityfixedh}
\mathbf{S}(p_1,p_2;h,a)\simeq \mathbf{S}^{-1}(-p_1,-p_2;h,-a)\, .
\end{equation}
These properties look like $q$-deformed generalizations of the parity transformation property of the $\ads$ world-sheet S-matrix, see e.g. \cite{Arutyunov:2009ga}, but have little to do with actual world-sheet parity as they do not leave $q$ invariant. The relevant behaviour of our S-matrix with regard to parity is
\begin{equation}
\label{eq:Sparityfixedh}
\mathbf{S}(-p_1,-p_2;h,a)\simeq \left(B\otimes B\right)\,  \mathbf{S}^{-1}(p_1,p_2;h,a) \, \left(B\otimes B\right)^{-1}\, ,
\end{equation}
where $B= A\otimes A$ with $A= \mbox{diag}(\sigma_1,\sigma_1)$, and $\sigma_1$ is the first Pauli matrix.\footnote{Not coincidentally, this is precisely the similarity transformation involved in the pseudo-unitarity of the phase-deformed S-matrix \cite{qTBAI,StijnsThesis}.} This shows that the spectrum of our theory is parity invariant. The matching behaviour of the dressing phase under the above transformations is discussed in appendix \ref{app:dressingphase}.

\subsection*{The dispersion relation}

With a proposal for the exact world-sheet S-matrix in hand, the other ingredient we need is the dispersion relation. To get the world-sheet dispersion relations we extend the semi-classics--compatible identification \eqref{eq:Eandpdef} to the full theory, and via eqn. \eqref{eq:shortening} immediately obtain
\begin{equation}
\label{eq:dispersionconstraint}
\cos^2\frac{\r}{2} \sinh^2 \frac{a \,\mathcal{E}}{2} - \sin^2\frac{\r}{2}\sin^2\frac{p}{2}=\sinh^2 \frac{a}{2}\, ,
\end{equation}
or just
\begin{equation}
\label{eq:dispersionsolved}
\mathcal{E}(p) = \frac{2}{a} \arcsinh \sqrt{\sec^2\frac{\r}{2}\sinh^2 \frac{a}{2} + \tan^2\frac{\r}{2} \sin^2 \frac{p}{2}}\,
\end{equation}
for the positive energy branch. Note that this dispersion relation is invariant under changes of sign of $a$ and $\r$. It is clearly not relativistic, and is not invariant under the mirror transformation
\begin{equation}
\label{eq:mirrortfEandp}
E \rightarrow i \tilde{p} \, , \,\,\,\,\,\,\, p\rightarrow i \tilde{E}\,,
\end{equation}
where $\tilde{p}$ and $\tilde{E}$ are the mirror momentum and energy, respectively.
This dispersion relation does exhibit an interesting feature however. Namely, if we take eqn. \eqref{eq:dispersionconstraint} and combine a double Wick rotation with a rescaling of the energy and momentum as $\mathcal{E}\rightarrow \pm \mathcal{E}/a$ and $p\rightarrow \pm p \,a$ (all choices of signs), the dispersion relation at $\r=\r_0$ becomes precisely that of our model at $\r=\r_0+\pi$, \emph{without} the Wick rotations! This is very interesting as it suggests a relation between world-sheet theories with $\r \in (-\pi/2,\pi/2]$ and the mirror versions of world-sheet theories with $\r \in (-\pi,-\pi/2] \cup (\pi/2,\pi]$, and vice versa. Let us discuss this relation in more detail.

\section{Mirror models and mirror duality}
\label{sec:mirrorduality}

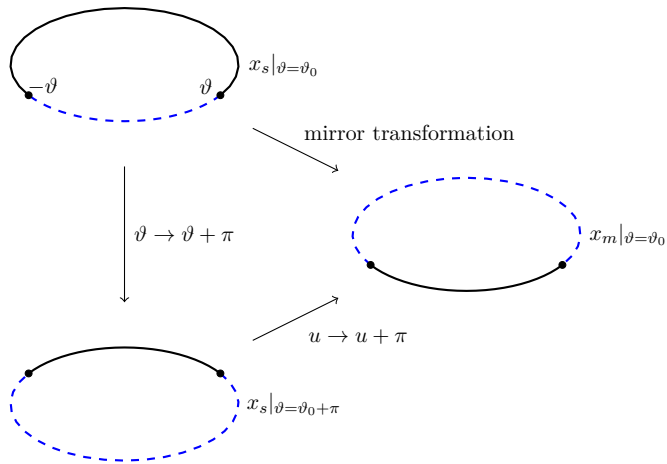
\begin{figure}
\begin{center}
\begin{tikzpicture}[scale=0.75, every node/.style={scale=0.75}]
\draw[black, thick, domain=-pi/2-4/7:pi/2+4/7] plot ({2*sin(\x r)-3},{cos(\x r)});
\draw[blue, dashed, thick, domain=pi/2+4/7:3*pi/2-4/7] plot ({2*sin(\x r)-3},{cos(\x r)});
\draw[blue, dashed, thick, domain=-pi/2-4/7:pi/2+4/7] plot ({2*sin(\x r)+3},{cos(\x r)-3});
\draw[black, thick, domain=pi/2+4/7:3*pi/2-4/7] plot ({2*sin(\x r)+3},{cos(\x r)-3});
\draw[black, thick, domain=-pi/2+4/7:pi/2-4/7] plot ({2*sin(\x r)-3},{cos(\x r)-6});
\draw[blue, dashed, thick, domain=pi/2-4/7:3*pi/2+4/7] plot ({2*sin(\x r)-3},{cos(\x r)-6});
\draw [->] (-3,-1.8) -- (-3,-4.2);
\draw [->] (-6/8,-3/2+3/8)-- (6/8,-3/2-3/8);
\draw [->] (-6/8,-3/2-3/8-3)-- (6/8,-3/2+3/8-3);
\node at (2,-3/2+3/10) {mirror transformation};
\node at (-3+1.05,-3) {$\r\rightarrow \r+\pi$};
\node at (1.1,-3/2-3/10-3) {$u\rightarrow u+\pi$};
\node at (-1/2+0.3,0) {$x_s|_{\r=\r_0}$};
\node at (3+3-1/2+0.34,-3) {$x_m|_{\r=\r_0}$};
\node at (-1/2+0.47,-6) {$x_s|_{\r=\r_0+\pi}$};
\node at ({2*sin(-(pi/2+4/7) r)-3+0.25},{cos(-(pi/2+4/7) r)+0.15}){$-\r$};
\node at ({2*sin((pi/2+4/7) r)-3-0.25},{cos((pi/2+4/7) r)+0.15}){$\r$};
\draw[fill] ({2*sin((pi/2+4/7) r)-3},{cos((pi/2+4/7) r)}) circle [radius=0.06];
\draw[fill] ({2*sin(-(pi/2+4/7) r)-3},{cos(-(pi/2+4/7) r)}) circle [radius=0.06];
\draw[fill] ({2*sin((pi/2+4/7) r)+3},{cos((pi/2+4/7) r)-3}) circle [radius=0.06];
\draw[fill] ({2*sin(-(pi/2+4/7) r)+3},{cos(-(pi/2+4/7) r)-3}) circle [radius=0.06];
\draw[fill] ({2*sin((-pi/2+4/7) r)-3},{cos((-pi/2+4/7) r)-6}) circle [radius=0.06];
\draw[fill] ({2*sin(-(3*pi/2+4/7) r)-3},{cos(-(3*pi/2+4/7) r)-6}) circle [radius=0.06];
\end{tikzpicture}
\end{center}
\caption{Mirror duality for the $x$-functions. The cuts of the $x_s$ function with $\r$ shifted by $\pi$ (bottom) are of the same length as the original $x_m$ function (right). In fact, the corresponding functions are identical upon shifting $u$ (rotating) by $\pi$.}
\label{fig:cutduality}
\end{figure}

We can parametrize all real values of momentum and (positive) energy satisfying our deformed dispersion relation \eqref{eq:dispersionsolved} by a real rapidity $u$ by identifying the $x$-functions in eqs. \eqref{eq:Eandpdef} by the $x_s$-function \eqref{eq:xs}. Requiring the mirror transformation \eqref{eq:mirrortfEandp}  to result in positive mirror energies and real mirror momenta we find that this is accomplished by the analytic continuation\footnote{This can of course be discussed on the uniformizing torus (see \cite{qTBAI} or \cite{StijnsThesis}), here we opted to keep these technical details to a minimum. Also, note that the transformation requires an inverse on $x_m$ should we wish to consider $a<0$ in our current parametrization.}
\begin{equation}
x_s \rightarrow x_m \,.
\end{equation}
Let us stress that this mirror transformation relates a model at a given value of $\r$ to its mirrored cousin at the \emph{same} value of $\r$. The resulting values of the central charges, denoted $\tilde{U}$ and $\tilde{V}$, are related to the mirror energy and momentum via eqn. \eqref{eq:Eandpdef} coupled with the mirror transformation \eqref{eq:mirrortfEandp}, or just
\begin{equation}
\label{eq:mirrorEandpfromUandV}
\tilde{V}^{2} = q^{i \tilde{p}}\,, \,\,\, \mbox{and} \,\,\,\, \tilde{U}^{2} = e^{-\tilde{E}}\, .
\end{equation}
Now we saw that at the level of the dispersion relation (shortening condition) shifting $\r$ by $\pi$ was somehow closely related to the mirror transformation. In fact, shifting $u$ in the same manner, at this level it is \emph{equivalent} to the mirror transformation! We have illustrated this idea in figure \ref{fig:cutduality}.
Concretely we have\footnote{In line with figure \ref{fig:cutduality}, note that since our rapidity is a periodic variable, branch points at $\pm \r$ are equivalent, or `dual', to branch points at $\pm (\r+\pi)$, only our way of connecting the branch cut through the origin or $\pi$ distinguishes our string and mirror $x$-functions. Shifting the rapidity by $\pi$, the branch cuts of the string and mirror functions coincide with those of the mirror and string functions at the dual $\r$ value respectively. The rest is a manner of conventions.}
\begin{equation}
\label{eq:xfunctionduality}
\left. x_{s}(u+\pi)\right|_{\r=\r_0+\pi} = \left.x_{m}(u)\right|_{\r=\r_0}\, ,
\end{equation}
anywhere on the cylinder with cuts. Since we will be considering these shifts often, let us introduce the notation
\begin{equation}
\label{eq:dualityshiftdef}
\left.f(\{u\})\right|_{\rd_0} \equiv \left. f(\{u+\pi\})\right|_{\r=\r_0+\pi} \, , \,\,\, \left.f(\{u\})\right|_{\r_0} \equiv \left. f(\{u\})\right|_{\r=\r_0} \,.
\end{equation}
Noting that $\xi\rightarrow 1/\xi$, under this `duality' transformation we then have\begin{equation}
e^{ip} = U^2 \rightarrow \tilde{V}^{2} = q^{i \tilde{p}}\,, \,\,\,\, q^{E} = V^2 \rightarrow \tilde{U}^{2} = e^{-\tilde{E}}\, ,
\end{equation}
as readily follows from the shortening condition rewritten as
\bea
\label{eq:impshort}
q^2\frac{x^+}{x^-}\frac{x^- +\xi}{x^+ +\xi}\frac{x^- +1/\xi}{x^+ +1/\xi}=1\, .
\eea
In other words we have
\begin{equation}
\label{eq:Eandpduality}
\left. a \, E(u)\right|_{\rd_0} =  \tilde{E}(u)|_{\r_0} \, , \,\,\,\, \mbox{while} \,\,\, \left. p(u)\right|_{\rd_0} = \left. -a \, \tilde{p}(u)\right|_{\r_0} .
\end{equation}
The relative sign in the relation between the momenta will be important later. Before moving on, note that the parity transformations $p \rightarrow -p$ and $\tilde{p} \rightarrow -\tilde{p}$ both correspond to $u\rightarrow -u$, with the $x$-functions satisfying
\begin{equation}
\label{eq:xfunctionparity}
x_s(-u)  = - \frac{x_s(u)+\xi}{x_s(u)\,\xi  +1} \, , \,\,\mbox{ and }\,\,\, x_m(-u)  = - \frac{ x_m(u)\, \xi  +1}{x_m(u)+\xi}
\end{equation}
Moving our discussion beyond the dispersion relation and on to the scattering theory is a little more involved but can be readily done.

\subsection*{Mirror duality of the scattering theory}

We can understand how to relate the scattering theory of the dualized string theory to that of the mirror theory through their respective bound state structures. As is well known from the undeformed case, the bound state pole of the string S-matrix in the bosonic channel (formally) corresponds to a zero in the fermionic channel of the mirror theory, and vice versa. This means that under mirror duality we need bosonic zeroes and poles to turn into fermionic poles and zeroes respectively. Since the notion of pole and zero in this context exchange under a change of sign of momentum (recall relation \eqref{eq:Eandpduality}), the simplest way this can be realized is for mirror duality to relate string and mirror S-matrices by flipping the sign of momentum and interchanging bosons and fermions. Indeed, upon taking the grading into account\footnote{See e.g. section 3.1.2 of \cite{Arutyunov:2009ga}. At the level of the graded S-matrix we really have a similarity transformation by a graded tensor product of $M$s.} this is precisely what happens, namely
\begin{equation}
\label{eq:Smatrixmirrorduality}
\big(D \,\check{\otimes}\, D\big) \, \mathbf{S}(p_1,p_2)\,\big(\hat{D} \,\check{\otimes}\, \hat{D}\big)|_{\rd_0}  =\tilde{\mathbf{S}}(-\tilde{p}_1,-\tilde{p}_2)|_{\r_0}\, ,
\end{equation}
where $D=(-1)^F M\otimes M$ and $\hat{D}= M\otimes(-1)^F  M$, $F$ is the fermion number operator in $\mathbb{C}^{2|2}$, and $M$ is the matrix representation of the permutation $(3412)$ of the basis of $\mathbb{C}^{2|2}$. We can get eqn. \eqref{eq:Smatrixmirrorduality} as follows. Firstly, the matrix part of the S-matrix simply flips the signs of its momenta under the action of $D$, up to the fermionic scattering amplitude $a_3$ given in appendix \ref{app:qdefmatrixSmatrix}, \emph{i.e.}
\begin{equation}
\label{eq:Smatrixbosonfermioninversion}
D\, S(p_1,p_2)\, \hat{D} = a_3(p_1,p_2)\, S(-p_1,-p_2)\, .
\end{equation}
This leaves the scalar factors to be matched. Let us recall
\begin{equation}
\label{eq:Ssu2defindualitysection}
S_{\mathfrak{su}(2)}(p_1,p_2)=\frac{1}{\sigma^2(p_1,p_2)}\frac{x_1^+ + \xi}{x_1^- + \xi}\frac{x_2^- + \xi}{x_2^+ + \xi}\cdot
\frac{x_1^- - x_2^+}{x_1^+ - x_2^-}\frac{1-\frac{1}{x_1^-x_2^+}}{1-\frac{1}{x_1^+x_2^-}}\, ,
\end{equation}
and introduce the natural mirror scalar factor
\begin{equation}
\label{eq:Ssl2def}
S_{\mathfrak{sl}(2)}(\tilde{p}_1,\tilde{p}_2) =\frac{1}{\Sigma^2(\tilde{p}_1,\tilde{p}_2)}\frac{x_1^-}{x_1^+}\frac{x_2^+}{x_2^-}\frac{x_1^+ + \xi}{x_1^- + \xi}\frac{x_2^- + \xi}{x_2^+ + \xi}
\frac{x_1^+ - x_2^-}{x_1^- - x_2^+}\frac{1-\frac{1}{x_1^+x_2^-}}{1-\frac{1}{x_1^-x_2^+}}\, ,
\end{equation}
where
\begin{equation}
\Sigma(\tilde{p}_1,\tilde{p}_2) \equiv \frac{1-\frac{1}{x_1^+ x_2^-}}{1-\frac{1}{x_1^- x_2^+}}\sigma(\tilde{p}_1,\tilde{p}_2)\, ,
\end{equation}
is the improved mirror dressing phase. We will always signify analytic continuation to the mirror theory by explicit dependence on $\tilde{p}$. Formally $S_{\mathfrak{sl}(2)}$ is just the bosonic scalar scattering amplitude $S_{\mathfrak{su}(2)}$ times the fermionic matrix scattering amplitude $(a_3)^2$. Now eqn. \eqref{eq:Smatrixmirrorduality} follows from two copies of eqn. \eqref{eq:Smatrixbosonfermioninversion} multiplied by $S_{\mathfrak{su}(2)}$, provided
\begin{equation}
a_3^2(p_1,p_2) \, S_{\mathfrak{su}(2)}(p_1,p_2)|_{\rd_0} = S_{\mathfrak{su}(2)}(-\tilde{p}_1,-\tilde{p}_2)|_{\r_0}\, ,
\end{equation}
or, in terms of the natural quantities,
\begin{equation}
\label{eq:scalarfactormirrorduality}
S_{\mathfrak{su}(2)}(p_1,p_2)|_{\rd_0} = S_{\mathfrak{sl}(2)}(-\tilde{p}_1,-\tilde{p}_2)|_{\r_0}\, .
\end{equation}
Comparing expressions \eqref{eq:Ssu2defindualitysection} and \eqref{eq:Ssl2def} and taking into account eqs. (\ref{eq:impshort}) and \eqref{eq:xfunctionparity} and the fact that $\xi$ is inverted at $\r+\pi$, we readily see this holds provided we factor out the corresponding dressing phases
\begin{equation}
\left. \sigma^2(p_i,p_j) S_{\mathfrak{su}(2)}(p_i,p_j)\right|_{\rd_0} = \left.\Sigma^2(-\tilde{p}_i,-\tilde{p}_j) S_{\mathfrak{sl}(2)}(-\tilde{p}_i,-\tilde{p}_j)\right|_{\r_0} \, .
\end{equation}
The most nontrivial statement is then that
\begin{equation}
\label{eq:dressingduality}
\left.\sigma(p_i,p_j)\right|_{\rd_0} = \left.\Sigma(-\tilde{p_i},-\tilde{p_j})\right|_{\r_0}\,,
\end{equation}
which holds within the so-called physical strip $-ia<\mbox{Im}(u)<ia$, as can be readily verified numerically. Outside this region there is no immediate agreement because the cut structure of these two objects is complementary by construction, due to the shift of $u$ by $\pi$ under the duality (on the torus this is a shift by a quarter of the real period), but this is just a matter of analytic continuation. In line with this, the crossing equations are compatible when understood to have their dual crossing transformation implemented in the opposite direction on the torus.

In the above discussion we could have traded the sign on momenta for inverse S-matrices (cf. eqn. \eqref{eq:Sparityfixedh}), but we feel this might cloud the physical picture. Of course these statements of mirror duality concretely translate to the diagonalized level.

\subsection*{Mirror duality of the Bethe-Yang equations}

Let us start with the Bethe-Yang equations for our deformed string theory \cite{qTBAI,qTBAII}
\begin{align}
\label{eq:BYstring}
1 = e^{i p_i J} \prod_{i\neq k}^{K^{\mathrm{I}}} S_{\mathfrak{su}(2)}(p_i,p_k)\prod_{\alpha=l,r}\prod_{i=1}^{K^{\mathrm{II}}_{\alpha}} \frac{1}{\sqrt{q}}\frac{y_i^{(\alpha)} - x_k^+}{y_i^{(\alpha)} - x_k^-}\sqrt{\frac{x_k^-}{x_k^+}},
\end{align}
where cf. eqn. \eqref{eq:Ssu2def}
\begin{equation}
S_{\mathfrak{su}(2)}(p_1,p_2)=\frac{1}{\sigma^2(p_1,p_2)}\frac{x_1^+ + \xi}{x_1^- + \xi}\frac{x_2^- + \xi}{x_2^+ + \xi}\cdot
\frac{x_1^- - x_2^+}{x_1^+ - x_2^-}\frac{1-\frac{1}{x_1^-x_2^+}}{1-\frac{1}{x_1^+x_2^-}}\, ,
\end{equation}
along with a set of auxiliary Bethe equations for each $\alpha$
\begin{align}
\label{eq:auxBEy}
1&= \prod_{i=1}^{K^{\mathrm{I}}}\sqrt{q}\frac{y_k - x^-_i}{y_k - x^+_i}\sqrt{\frac{x^+_i}{x^-_i}}
\prod_{i=1}^{K^{\mathrm{III}}}\frac{\sin{\frac{1}{2}(\nu_k - w_i-i a)}}{\sin{\frac{1}{2}(\nu_k - w_i+i a)}}\, ,\\
\label{eq:auxBEw}
-1&= \prod_{i=1}^{K^{\mathrm{II}}} \frac{\sin{\frac{1}{2}(w_k - \nu_i + i a)}}{\sin{\frac{1}{2}(w_k - \nu_i - i a)}}\prod_{j=1}^{K^{\mathrm{III}}}\frac{\sin{\frac{1}{2}(w_k - w_j - 2 i a)}}{\sin{\frac{1}{2}(w_k - w_j + 2 i a)}}\, ,
\end{align}
that arise in the standard manner by diagonalization of the transfer matrix through the (algebraic) Bethe ansatz, here done with respect to a \emph{bosonic} reference state (vacuum). Here $K^{\mathrm{I}}$ represents the number of world-sheet excitations, and the full set of excitation numbers is directly related to the set of global charges of our string, as described at the end of appendix
\ref{app:qdefmatrixSmatrix}. As usual, $y$ is related to $\nu$ as $x$ is to $u$ in eqn. \eqref{eq:rapiditydef}, meaning we are free to take $y=(x_s(\nu))^{\pm1}$, as long as we allow for both options. Parity invariance of the above equations follows by the relations \eqref{eq:xfunctionparity}, so that the full set of Bethe-Yang equations inverts upon flipping the sign of all rapidities (including the $\nu$s and $w$s).

Upon dualization we would like these equations to be equivalent to the mirror Bethe-Yang equations
\begin{align}
\label{eq:BYmirror}
1 = e^{i \tilde{p}_i R} \prod_{i\neq k}^{\tilde{K}^{\mathrm{I}}} S_{\mathfrak{sl}(2)}(\tilde{p}_i,\tilde{p}_k)\prod_{\alpha=l,r}\prod_{i=1}^{\tilde{K}^{\mathrm{II}}_{\alpha}} \sqrt{q}\frac{y_i^{(\alpha)} - x_k^-}{y_i^{(\alpha)} - x_k^+}\sqrt{\frac{x_k^+}{x_k^-}}\, ,
\end{align}
where the auxiliary Bethe equations are formally the same as eqs. \eqref{eq:auxBEy} and \eqref{eq:auxBEw}, but with fermionic excitation numbers $\tilde{K}$. These are equal to the bosonic ones, with the exception that
\begin{equation}
\label{eq:KIIdualization}
K^{\mathrm{II}} = \tilde{K}^{\mathrm I} - \tilde{K}^{\mathrm{II}} + 2 \tilde{K}^{\mathrm{III}}\, .
\end{equation}

To see that the Bethe-Yang equations are dual, first note that the duality transformation \eqref{eq:dualityshiftdef} formally leaves the auxiliary equations invariant (shift also the $\nu$s and $w$s by $\pi$ and identify $y$ via $(x_m(\nu)|_{\r_0})^{\pm1}$ rather than $(x_s(\nu)|_{\rd_0})^{\pm1}$). To match individual states we of course need to identify
\begin{equation}
\label{eq:chargeduality}
\{ K \} = \{ \tilde{K} \}\, ,
\end{equation}
under mirror duality, which is nothing but the action of $D$ translated to the diagonalized level. Next, by eqn. \eqref{eq:scalarfactormirrorduality}, dualization precisely turns the scattering terms in eqs. \eqref{eq:BYstring} into the inverse of those in eqs. \eqref{eq:BYmirror} (by parity). Combining this with the duality relation \eqref{eq:Eandpduality} between the momenta, we find that the dualized string Bethe equations are nothing but the mirror Bethe-Yang equations at $\r = \r_0$ in inverse form, under the above identification of charges (excitation numbers) and the identification
\begin{equation}
\label{eq:lengthduality}
R = a J \, .
\end{equation}
Let us emphasize that the relation between the momenta in eqn. \eqref{eq:Eandpduality} is taken into account by this rescaling of length; the momentum identification itself goes through \emph{without} rescaling. Provided we now take into account the energy rescaling in eqn. \eqref{eq:Eandpduality} the energy spectra of these theories will also manifestly agree. This shows that our string theories at $\r=\r_0+\pi$ and their mirrored versions at $\r=\r_0$ have identical dispersion relations and scattering properties at any given $a$; they represent one and the same theory. We will revisit the length identification \eqref{eq:lengthduality} at a later stage.

One note regarding the physical parameter space is in order. Earlier we saw that the (semi-classical) sigma model is parametrized by positive imaginary $\xi$. The duality transformation as described above would take $\xi$ out of this domain. To get back, we can combine the transformation \eqref{eq:dualityshiftdef} with the sign change $\xi\to-\xi$, which leaves the model invariant (cf. eqn. \eqref{eq:Sxiinvariance}). Mirror duality then relates $\r$ and $\pi-\r$. This form of mirror duality would introduce an ungraceful (inconsequential) minus sign in the duality relation \eqref{eq:xfunctionduality}, but is more appropriate when considering the sigma model directly.

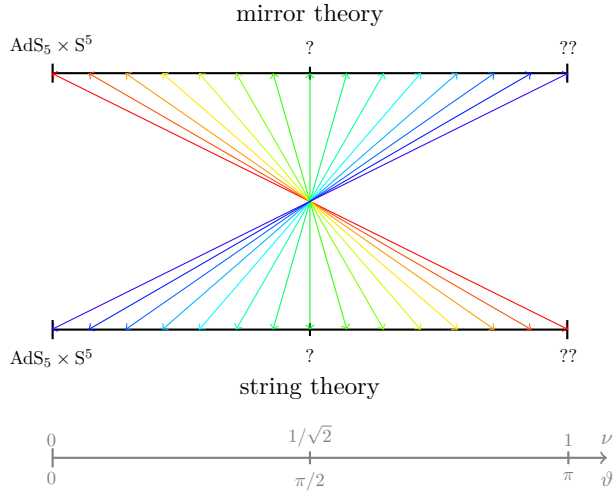
\begin{figure}
\begin{center}
\begin{tikzpicture}[scale=1.7, every node/.style={scale=0.85}]
\draw [thick][|-|] (-2,1) -- (2,1);
\draw [thick][|-|] (-2,-1) -- (2,-1);
\draw [thick] (0,-1.05) -- (0,-0.95);
\draw [thick] (0,1.05) -- (0,0.95);
\foreach \x in {0,0.05,...,0.71} {
	\definecolor{currentcolor}{hsb}{\x, 1, 1}
	\draw[draw=currentcolor][<->]
		({4*10/7*\x-2},1) -- ({-4*10/7*\x+2},-1);
}

\node [below] at (0,-1.3) {string theory};
\node [above] at (0,1.3) {mirror theory};

\draw [white!50!black,thick][|->] (-2,-1-0.9-0.1) -- (2+0.3,-1-0.9-0.1);
\draw [white!50!black,thick] (0,-1.95-0.1) -- (0,-1.85-0.1);
\draw [white!50!black,thick] (2,-1.95-0.1) -- (2,-1.85-0.1);
\node [white!50!black,below,scale=0.85] at (2.3,-1.95-0.1) {$\r$};
\node [white!50!black,below,scale=0.75] at (-2,-1.96-0.1) {$0$};
\node [white!50!black,below,scale=0.75] at (0,-1.96-0.1) {$\pi/2$};
\node [white!50!black,below,scale=0.75] at (2,-1.96-0.1) {$\pi$};
\node [white!50!black,above,scale=0.85] at (2.3,-1.85-0.1) {$\nu$};
\node [white!50!black,above,scale=0.75] at (-2,-1.86-0.1) {$0$};
\node [white!50!black,above,scale=0.75] at (0,-1.86-0.1) {$1/\sqrt{2}$};
\node [white!50!black,above,scale=0.75] at (2,-1.86-0.1) {$1$};

\node [below,scale=0.75] at (-2,-1.1) {$\ads$};
\node [below,scale=0.75] at (0,-1.1) {?};
\node [below,scale=0.75] at (2,-1.1) {??};
\node [above,scale=0.75] at (-2,1.1) {$\ads$};
\node [above,scale=0.75] at (0,1.1) {?};
\node [above,scale=0.75] at (2,1.1) {??};
\end{tikzpicture}
\end{center}
\caption{Mirror duality of the deformed light-cone string. Mirror duality states that the deformed string theories (bottom line) labeled by $\r$ are equivalent to double Wick rotated versions of these same string theories at $\pi-\r$. This equivalence in particular singles out the special cases at $\r=0, \pi/2$ and $\pi$ deserving particular investigation.}
\label{fig:physicalmirrorduality}
\end{figure}

It is worth emphasizing that from the point of view of mirror duality there are three distinguished models: the undeformed string at $\r=0$, a self-dual model at $\r=\pi/2$, and the `maximally' deformed model at $\r=\pi$. However, the latter is currently not well-defined as a sigma model. This is related to the fact that if we want to have sensible representations (finite $x^\pm$) in the limits $\r\rightarrow 0$ and $\r\rightarrow \pi$ we need to simultaneously take $q\rightarrow1$ ($a\rightarrow0$). For $\r \rightarrow 0$ this directly gives the undeformed model as indicated below eqn. \eqref{eq:xsmconjugation}. For $\r \rightarrow \pi$, we can take a similar sensible limit at the level of the $x$-functions. Taking $x_s(u g a)$ with $\r = 2 \arccos g \sinh a$ (instead of eqn. \eqref{eq:thetadef}) gives the undeformed $x_m$ function and vice versa in the limit $a\rightarrow0$. To get sensible central charges we need to rescale the energy and momentum by $a$ however, cf. relation \eqref{eq:Eandpduality}. In this way the integrable model at $\r =\pi$ precisely becomes the \emph{undeformed} $\ads$ \emph{mirror} model. However this limit is clearly singular in light of the length rescaling \eqref{eq:lengthduality} and the required rescaling of energy to match spectra. This matches the fact that without further field redefinitions the limit $\r\rightarrow \pi$ of the sigma model is singular, so that it could not be directly related to the finite $\ads$ mirror model. Still, since it is possible to extract our sensible undeformed mirror model out of the overarching integrable model description in the limit $\r \rightarrow \pi$, it would be very interesting to see to what extent this can be translated to the sigma model. In general, mirror duality of the integrable model translates to the physical equivalence illustrated in figure \ref{fig:physicalmirrorduality}. Let us now try to check these statements at the level of the sigma model.

\section{Mirror duality of the sigma model}
\label{sec:sigmamodelduality}

Demonstrating mirror duality directly for the full light-cone quantum sigma model is beyond our current reach. We can however provide some evidence for mirror duality of the sigma model at the semi-classical level. Along the way we will end up with the necessary ingredients to consider giant magnon solutions on our $\eta$-deformed background, which we will use to verify the dispersion relations of the string and mirror theories.

Let us recall that in the large $g$ limit $\xi$ tends to $i\frac{\nu}{\sqrt{1-\nu^2}} \equiv i\varkappa$, so that in the semiclassical regime the duality transformation $\xi\to -1/\xi$ is equivalent to $\varkappa\to 1/\varkappa$. Since semi-classically $a=-\tfrac{\varkappa}{g\sqrt{1+\varkappa^2}}$ and is to be held fixed under mirror duality, we should simultaneously rescale $g\to g/\varkappa$. In short, in the semiclassical regime the duality transformation of the two independent parameters of the sigma-model boils down to\footnote{Note that the self-dual point $\r=\pi/2$ corresponds semi-classically to $\varkappa = 1$, leaving $g$ invariant.}
\bea
\label{semiduality}
\varkappa\to \frac{1}{\varkappa}\, , ~~~~g\to \frac{g}{\varkappa}\, .
\eea
According to our previous discussion, this transformation of the original parameters of the string sigma-model should be equivalent to a double Wick rotation of the world-sheet coordinates
\bea
\label{dw}
\tau\to i\tilde{\sigma}\, , ~~~~~\sigma\to -i\tilde{\tau}\, ,
\eea
which relates the original and mirror theory.

To study the duality transformation in the world-sheet approach, we fix a uniform light-cone gauge depending on an auxiliary parameter $\agauge$ and find the corresponding gauge-fixed action. This is done in appendix \ref{app:GFA}. As in the undeformed case, the most simple expression for the gauge-fixed action is obtained in the gauge $\agauge=1$ and it reads
\bea
\label{Gac}
S=\int{\rm d}\tau{\rm d}\sigma\Bigg[-1+\sqrt{\frac{G_{\phi\phi}}{G_{tt}} \mathscr{X}}
+ \dot{x}^{\mu}x'^{\nu}B_{\mu\nu}\Bigg]\, ,\eea
where
\bea
\label{chi}
{\mathscr X}=1+G_{\mu\nu}\Big(\frac{1}{G_{\phi\phi}}\dot{x}^{\mu}\dot{x}^{\nu}&-&G_{tt}x'^{\mu}x'^{\nu}\Big)
-\frac{G_{tt}}{G_{\phi\phi}}\Big(G_{\mu\nu}G_{\tau\sigma}-G_{\mu\tau}G_{\nu\sigma}\Big)
\dot{x}^{\mu}\dot{x}^{\nu}x'^{\tau}x'^{\sigma}\, , ~~
\eea
and $x^{\mu}$ with $\mu=1,\ldots 8$ are eight transverse physical fields parametrizing the $\eta$-deformed background. The components of the metric as well as the Wess-Zumino term (B-field) entering eqs. \eqref{Gac} and \eqref{chi} are given at the end of appendix \ref{app:GFA}. The action (\ref{Gac}) is rather involved because of the complexity of the metric. Therefore we will restrict our attention to consistent reductions of (\ref{Gac}) which are given by switching off all the fields except one either on the deformed sphere or on deformed AdS. Later we will use the corresponding reduced actions to construct giant magnon solutions.

In general it appears that the sphere metric components $G_{ii}^{\alg{s}}$ look different for different $i$, {\it cf.} appendix \ref{app:GFA}.  Nonetheless, explicit calculation reveals that the resulting component $G_{ii}^{\alg{s}}$ is the same for any $i$ if we consider a fixed $i$ and switch off all the sphere and AdS fields except $y^i$. This shows that there exists a unique giant magnon. An analogous situation holds for the AdS part; the form of the AdS metric components $G_{ii}^{\alg{a}}$ does not depend on index $i$ provided we only keep the AdS field $z^i$  non-vanishing.

Keeping only a single field from the sphere non-vanishing, for instance $y^1$, and making a
change of variables $\mathpzc{y}=y^1/(1+(y^1)^2/4)$, we find that the action \eqref{Gac} reduces to
{\small
\begin{equation}
\label{Ssphere}
\hspace{-0.5cm}S^{\alg s}[g,\varkappa]=g\int {\rm d}\tau{\rm d}\sigma \, \Bigg[-1
+\sqrt{\frac{(1+\varkappa^2 {\mathpzc y}^2)\dot{\mathpzc y}^2-(1+\varkappa^2)(1-{\mathpzc y}^2){\mathpzc y}'^2+(1-{\mathpzc y}^2)^2(1+\varkappa^2 {\mathpzc y}^2)}
{(1-{\mathpzc y}^2)(1+\varkappa^2{\mathpzc y}^2)^2}}\Bigg]\,.
\end{equation}}

\noindent
If instead we choose $z^1$ to be non-vanishing and use the variable $\mathpzc{z}=z^1/(1-(z^1)^2/4)$, we find
{\small
\begin{equation}
\label{Sads}
\hspace{-0.4cm}S^{\alg a}[g,\varkappa]=g\int{\rm d}\tau{\rm d}\sigma \, \Bigg[-1
+\sqrt{\frac{(1-\varkappa^2 {\mathpzc z}^2)\dot{\mathpzc z}^2-(1+\varkappa^2)(1+{\mathpzc z}^2){\mathpzc z}'^2+(1+{\mathpzc z}^2)(1-\varkappa^2 {\mathpzc z}^2)^2}
{(1+{\mathpzc z}^2)^2(1-\varkappa^2{\mathpzc z}^2)}}\Bigg]\, .
\end{equation}
}

\noindent
We know from the undeformed string that the giant magnon solutions of the mirror theory arise from the AdS part of the corresponding action \cite{Arutyunov:2007tc}. With this in mind, we apply the double Wick rotation (\ref{dw}) to (\ref{Sads}) and get the following action for the mirror theory
{\small
\begin{equation}
\label{Sadsm}
\hspace{-0.4cm}\tilde{S}^{\alg a}[g,\varkappa]=g \int {\rm d}\tilde{\tau}{\rm d}\tilde{\sigma} \, \Bigg[-1
+\sqrt{\frac{(1+\varkappa^2)(1+{\mathpzc z}^2)\dot{{\mathpzc z}}^2-(1-\varkappa^2 {\mathpzc z}^2){\mathpzc z}'^2  +(1+{\mathpzc z}^2)(1-\varkappa^2 {\mathpzc z}^2)^2}
{(1+{\mathpzc z}^2)^2(1-\varkappa^2{\mathpzc z}^2)}}\Bigg]\, .
\end{equation}
}

\noindent
Now we are ready to compare the actions (\ref{Ssphere}) and (\ref{Sadsm}). First we apply the duality transformation (\ref{semiduality}) to the action (\ref{Ssphere}) which gives
{\small
\begin{align}
\label{Sspheredual}
S^{\alg s}[&g/\varkappa,1/\varkappa]=\\
\nonumber
&=\frac{g}{\varkappa}\int {\rm d}\tau{\rm d}\sigma \, \Bigg[-1
+\sqrt{\frac{(1+\varkappa^{-2} {\mathpzc y}^2)\dot{\mathpzc y}^2-(1+\varkappa^{-2})(1-{\mathpzc y}^2){\mathpzc y}'^2+(1-{\mathpzc y}^2)^2(1+\varkappa^{-2} {\mathpzc y}^2)}
{(1-{\mathpzc y}^2)(1+\varkappa^{-2}{\mathpzc y}^2)^2}}\Bigg].
\end{align}
}

\noindent
Rescaling now the field $\mathpzc{y}=\varkappa \mathpzc{u}$ and performing
a change of variables
\bea
\label{cv}
\tau\to \frac{\varkappa}{\sqrt{1+\varkappa^2}}\, \tilde{\tau}\,  , ~~~~~
\sigma\to \sqrt{1+\varkappa^2}\, \tilde{\sigma} \, ,
\eea
we find that
\bea
S^{\alg s}[g/\varkappa,1/\varkappa]=
\tilde{S}^{\alg a}[g,\varkappa]\, ,
\eea
provided we identify the field $\mathpzc{u}$ with $\mathpzc{z}$. The same relation holds for the single field actions $S^{\alg a}$ and $\tilde{S}^{\alg s}$. Note that the interchange of the AdS and sphere actions under our mirror duality is precisely what is reflected at the level of the Bethe ansatz by passing from the `$\mathfrak{su}(2)$'- to the `$\mathfrak{sl}(2)$'-grading. To match the rescalings \eqref{cv} with the energy and length rescalings \eqref{eq:Eandpduality} and \eqref{eq:lengthduality} respectively, note that the mirror time appropriate for the integrable model is $g\,\tilde{\tau}$ rather than $\tilde{\tau}$, and that in the above the spatial extent of the string was implicitly rescaled to $J/g$ (cf. eqn. \eqref{gfaction}), which gives a factor of $\varkappa/g$ after dualization. The analysis of these reduced actions suggests that in general we might expect
\bea
S[g/\varkappa,1/\varkappa]=
\tilde{S}[g,\varkappa]\,
\eea
as the manifestation of mirror duality in the semi-classical regime. It is an interesting open problem how to prove this statement for the whole model. Its non-triviality lies in finding proper field redefinitions which would allow one to identify the dualized action with the mirror one.
Now let us use these reduced actions to construct giant magnon solutions in both the original and the mirror theory.

\subsection*{Giant magnon}

To find a giant magnon solution \cite{Hofman:2006xt}, we consider the action (\ref{Ssphere}) and substitute a soliton ansatz
$$
{\mathpzc{y}}=Q(\sigma-v\tau)\, ,
$$
where $v$ is the velocity of the soliton.  This defines a reduced Lagrangian $L_{red}$ for a one-dimensional system where $\mathpzc{y}'$ plays the role of particle velocity. The corresponding conjugate momentum is $\pi=\frac{\pa L}{\pa \mathpzc{y}'}$. The reduced Hamiltonian is then
\bea
H_{red}=\pi \mathpzc{y}'-L_{red}\, .
\eea
Fixing the level $H_{red}=0$ corresponding to motion in an infinite volume, we find the differential equation
\bea
Q'=\frac{\sqrt{1+\varkappa^2}\, Q(1-Q^2)}{\sqrt{1-v^2+\varkappa^2 -(1+(1+v^2)\varkappa^2)Q^2}}\, ,
\eea
governing the soliton profile. We assume that $|v|<1$ and that the motion happens between two turning points
\begin{equation*}
0 < Q <Q_{max}\, , ~~~~Q_{max}=\sqrt{\frac{1+\varkappa^2-v^2}{1+\varkappa^2+\varkappa^2v^2}}<1\, .
\end{equation*}
The density of the world-sheet momentum is $\mathpzc{p}=\frac{\pa \mathscr {L}}{\pa \dot{y}}$ and on the soliton solution it reduces to
\bea
\mathpzc{p}_{\mathpzc{y}}=-\frac{\sqrt{1+\varkappa^2}\, Qv}{(1-Q^2)\sqrt{1-v^2+\varkappa^2 -(1+(1+v^2)\varkappa^2)Q^2}}\, .
\eea
This gives rise to a total world-sheet momentum of
\bea
p=-2\int_{-\infty}^{\infty}{\rm d}\sigma \mathpzc{p}_{\mathpzc{y}}\mathpzc{y}'=2\int_0^{Q_{max}}{\rm d}Q\, |\mathpzc{p}_{\mathpzc{y}}|
=2\, {\rm arccot}\, \frac{v\sqrt{1+\varkappa^2}}{\sqrt{1+\varkappa^2-v^2}}\, .
\eea
The energy of the soliton is obtained by integrating the Hamiltonian density
${\mathcal H}= \mathpzc{p}_{\mathpzc{y}}\dot{\mathpzc{y}}-\mathscr{L}$
which gives
\bea
E=g\int_{-\infty}^{\infty} {\rm d}\sigma{\mathcal H}=2g\int_0^{Q_{max}}{\rm d}Q \frac{\mathcal{H}}{|Q'|}=\frac{2g\sqrt{1+\varkappa^2}}{\varkappa}
{\rm arctanh}\frac{\varkappa\sqrt{1+\varkappa^2-v^2}}{1+\varkappa^2}\, .
\eea
Solving for $v$ in terms of $p$ and substituting the answer into the expression for energy, we find
\bea
E=\frac{2g\sqrt{1+\varkappa^2}}{\varkappa}{\rm arcsinh} \, \Big|\varkappa\sin\frac{p}{2}\Big|=\frac{2g}{\nu}
{\rm arcsinh} \, \frac{\nu}{\sqrt{1-\nu^2}}\Big|\sin\frac{p}{2}\Big|\, , \eea
which is precisely the large $g$ limit of the dispersion relation \eqref{eq:dispersionsolved}.

\subsection*{Mirror giant magnon}

To obtain the mirror giant magnon we can repeat the above computation, starting instead from the action (\ref{Sadsm}) and ansatz
$$
{\mathpzc{z}}={\mathcal Q}(\sigma-v\tau)\,
$$
In this case the differential equation for the soliton profile corresponding to the zero energy level of the reduced Hamiltonian is
\bea
{\mathcal Q}'=\frac{\sqrt{1+\varkappa^2}\, {\mathcal Q}(1-\varkappa^2 {\mathcal Q}^2)}
{\sqrt{1-v^2(1+\varkappa^2)-(\varkappa^2+(1+\varkappa^2)v^2){\mathcal Q}^2}}\, .
\eea
The motion happens between two turning points
$$
0 < {\mathcal Q}<{\mathcal Q}_{max}\, , ~~~~{\mathcal Q}_{max}=\sqrt{\frac{1-(1+\varkappa^2)v^2}{\varkappa^2+(1+\varkappa^2)v^2}}<\frac{1}{\varkappa}\, .
$$
The total world-sheet momentum of the mirror soliton is
\bea
\tilde{p}=\frac{2\sqrt{1+\varkappa^2}}{\varkappa}\arctan\Bigg(\frac{\varkappa\sqrt{1-(1+\varkappa^2)v^2}}{v(1+\varkappa^2)}\Bigg)\, ,
\eea
while for the energy we obtain
\bea
\tilde{E}=2g\, {\rm arcsinh}\sqrt{\frac{1-(1+\varkappa^2)v^2}{\varkappa^2+(1+\varkappa^2)v^2}}\, .
\eea
Expressing $\tilde{E}$ in terms of $\tilde{p}$, we find
\bea
\tilde{E}=2g\, {\rm arcsinh} \Bigg|\frac{1}{\varkappa} \sin\frac{\varkappa }{\sqrt{1+\varkappa^2}}\frac{\tilde{p}}{2} \Bigg| =
2g \, {\rm arcsinh} \Bigg|\frac{\sqrt{1-\nu^2}}{\nu}\sin\frac{\nu \tilde{p}}{2}  \Bigg|
\, .
\eea
Upon rescaling $\tilde{\tau} \rightarrow \tilde{\tau}/g$ this dispersion relation matches the large $g$ limit of the mirror version of \eqref{eq:dispersionsolved} (keeping $\tilde{p}/g$ fixed there, analogous to the undeformed case \cite{Arutyunov:2007tc}).

\section{The Thermodynamic Bethe Ansatz}

\label{sec:TBA}
As in the undeformed case,
 the finite size ground state of the deformed string theories can be obtained from the free energy of their mirror models, and these free energies can be computed by means of the TBA. Deriving the corresponding TBA equations is a text-book procedure once the appropriate string hypothesis has been formulated (see e.g. \cite{EFGKK}). As we might expect on representation theoretical grounds, in the present case there is no qualitative change in the string hypothesis as we move from the $\ads$ mirror model to our current real-$q$--deformed mirror model. As such, our string hypothesis will be identical to the one for the $\ads$ superstring \cite{Arutyunov:2009zu}, up to parametrization of course. As a consequence, also the TBA equations are qualitatively identical to those of the undeformed string \cite{Gromov:2009bc,Bombardelli:2009ns,Arutyunov:2009ur}. The change in parametrization can be described very simply; all particle types that lived on the real rapidity line, now live on the real rapidity interval $(-\pi,\pi]$, and the $y$-particles that lived on the interval between the branch points of the undeformed $x$-functions at $\pm2$, now live between the branch points of the our $x$-functions at $\pm \r$.\footnote{This last fact follows by the usual arguments, using the conjugation property of $x_m$, parametrizing the resulting allowed values of $y$ (those being $|y|^2-1 = (y-y^*)\xi$) and substituting this in the definition of our rapidity, eqn. \eqref{eq:rapiditydef}. See also the discussion around eqs. (\ref{eq:dressingcontourdef}-\ref{eq:rapiditycontourparam2})}
At the level of the simplified TBA equations \cite{Arutyunov:2009ux} the deformation can be (almost completely) implemented in a particularly simple fashion. The deformation amounts to taking the convolutions to be
\begin{align}
f\star h(u,v)=\,&\int_{-\pi}^{\pi}\, dt\, f(u,t)h(t,v) \, , \\
f\, \hat{\star}\,  h(u,v)=\,&\int_{-\r}^{\r}\, dt\, f(u,t)h(t,v)\, , \\
f\, \check{\star}\, h(u,v)=\,&\int_{-\pi}^{-\r}\, dt\, f(u,t)h(t,v)+\int_{\r}^{\pi}\, dt\, f(u,t)h(t,v)\, ,
\end{align}
noting that all Y-functions now live on a $u$-cylinder (with cuts), and replacing the standard kernel $s(u) = (4 \cosh{\frac{\pi u}{2}})^{-1}$ by its doubly periodic analogue
\begin{equation}
\label{eq:sigmakerneldef}
s(u) \rightarrow \sum_{n \in \mathbb{Z}} s\left(\tfrac{u + 2\pi n}{a}\right) =  \sum_{n\in \mathbb{Z}} \frac{1}{4 a \cosh \frac{\pi (u+ 2\pi n)}{2 a}} \equiv \s(u)\, .
\end{equation}
Of course $\s$ is just a suitably normalized Jacobi elliptic $\dn$ function
\begin{equation}
\s(u) = \frac{K(m^\prime)}{2\pi a} \, \dn(u)\, ,
\end{equation}
with real period $2\pi$ and imaginary period $4 a$, where $K(m^\prime)$ is the elliptic integral of the corresponding complementary elliptic modulus.\footnote{To manifestly match the normalisations note that $K(m^\prime) \, = \sum_{l \in \mathbb{Z}} \frac{\pi}{2\cosh\pi^2 l/a}$.}

\subsection*{The equations}

The ground state energies of our deformed strings are given by
\bea
\label{eq:Energy} E(J) &=&-\int_{-\pi}^{\pi} du\, \sum_Q \frac{1}{2\pi}\frac{d\tilde{p}^Q}{du}\log\left(1+Y_Q\right)\,,
\eea
where $\tilde{p}^{Q}$ is the momentum of a $Q$ particle mirror bound state, defined in appendix \ref{app:Smatricesandkernels} along with the other kernels used below, and the Y-functions $Y_Q$ are solutions of the simplified TBA equations
\begin{align}
\log Y_Q = \, & \log{Y_{Q+1}Y_{Q-1}}\star \s + \log\left(1+\frac{1}{Y_{Q-1|vw}^{(1)}}\right)\left(1+\frac{1}{Y_{Q-1|vw}^{(2)}}\right)\\
& -\log (1+Y_{Q-1})(1+Y_{Q+1})\star \s \, ,\nonumber
\end{align}
for  $Q>1$, while for $Q=1$
\begin{equation}
\log Y_1=\sum_a \log\left(1-\tfrac{1}{Y_-^{(\alpha)}}\right)\, \hat{\star}\, \s - \log \left(1+\tfrac{1}{Y_2}\right) \star \s -\check{\Delta}\,\check{\star}\, \s\, ,
\end{equation}
where
\begin{align}
\check{\Delta}=&L\check{\cal E}+\sum_a \log \left(1-\tfrac{1}{Y_-^{(\alpha)}}\right)\left(1-\tfrac{1}{Y_+^{(\alpha)}}\right)\star \check{K}+2\log(1+Y_Q)\star
\check{K}^{\Sigma}_Q\\
\nonumber
&\quad\quad\quad +\sum_a \log\left(1-\tfrac{1}{Y_{M|vw}^{(\alpha)}}\right)\star \check{K}_M\, .
\end{align}
The $Y_\pm$ functions (for $y$-particles associated to $(x_m(u))^{\mp1}$) appearing above satisfy
\begin{align}
\log\frac{Y_{+}^{(\alpha)}}{Y_{-}^{(\alpha)}} = &
\log(1+Y_{Q})\star K_{Qy}\,,\\
\log Y_{-}^{(\alpha)}Y_{+}^{(\alpha)}  = & \,-\log(1+Y_{Q})\star K_{Q}+
2\log(1+Y_{Q})\star K_{xv}^{Q1}\star \s+2\log\frac{1+Y_{1|vw}^{
{(\alpha)}}}{1+Y_{1|w}^{(\alpha)}}\star \s\,, \nonumber
\end{align}
while the  $Y_{M|(v)w}$ functions satisfy
\begin{align}
\log{Y_{M|vw}^{(\alpha)}} = & \,\log{(1+Y_{M+1|vw}^{(\alpha)})(1+Y_{M-1|vw}^{(\alpha)})} \star  \s - \log(1+Y_{M+1})\star  \s \label{eq:TBAvwbulk}\\
& \hspace{20pt} + \delta_{M,1} \log{\left(\frac{1-Y_-^{(\alpha)}}{1-Y_+^{(\alpha)}}\right)}  \, \hat{\star} \, \s\, , \nonumber\\
\log{Y_{M|w}^{(\alpha)}} & = \log{(1+Y_{M+1|w}^{(\alpha)})(1+Y_{M-1|w}^{(\alpha)})} \star  \s + \delta_{M,1} \log{\left(\frac{1-\frac{1}{Y_-^{(\alpha)}}}{1-\frac{1}{Y_+^{(\alpha)}}}\right)}  \, \hat{\star} \, \s\, , \label{eq:TBAwbulk}
\end{align}
where $Y_{0|(v)w}=0$.

As usual we can define an operator $s^{-1}$, which now acts as
\begin{equation}
 f \circ s^{-1}(u) = \lim_{\epsilon\rightarrow 0} f(u + i a - i\epsilon) +  f(u - i a + i\epsilon)\, ,
\end{equation}
so that
\begin{equation}
(f \star \s) \circ s^{-1}(u) = f(u)\, , \, \, \, \, \mbox{for} \, \, u \in (-\pi,\pi] \, ,
\end{equation}
Applying $s^{-1}$ to the above TBA equations directly gives the Y-system.\footnote{As for the undeformed string, there is no Y-system equation for $Y_+$, and to get the Y-system equation for $Y_-$ we need the identities  $K^{Qy}_- \circ s^{-1} =  K_{xv}^{Q1} + \delta_{Q,1}$ and $K_M \circ s^{-1} = K_{M1} + \delta_{M,1} $.}

\subsection*{The ground state energy}

Since the full symmetry algebra of our deformed string theory is currently not known, we do not know whether the ground state is a protected state (indeed, in the undeformed case this relies on $\mathfrak{psu}(2,2|4)$ symmetry, not just the light-cone symmetry $\mathfrak{psu}(2|2)^{\oplus 2}$). Still, we can ask whether the ground state TBA equations have a simple and natural solution with zero energy, which is indeed the case. Since our kernels are identically normalized (on their appropriate intervals) to the corresponding undeformed kernels, and chemical potentials resulting from twisting enter the TBA equations \cite{Ahn:2011xq,deLeeuw:2012hp} (see also \cite{StijnsThesis}) identically as well, the considerations of \cite{Frolov:2009in} (see also \cite{deLeeuw:2012hp}) go through directly, and we conclude there exists a solution of the ground state TBA equations with all $Y_Q=0$, and hence zero energy, given by the same constant Y-functions as in the undeformed case. It thus appears that the ground state remains a protected state in our deformed theories.

In fact, we can now also take a look at the length duality \eqref{eq:lengthduality} from the TBA point of view. As discussed in \cite{Frolov:2009in}, in the undeformed case analyticity of the Y-functions in the TBA equations requires quantization of $J$ (the inverse mirror temperature), matching its origin as quantized angular momentum. This arises because the Y-functions contain a factor of $e^{-J \tilde{E}}$, which is only meromorphic on the torus for (half-)integer $J$. Though not argued in \cite{Frolov:2009in}, a TBA description of the thermodynamics of infinitely long ($J\rightarrow \infty$) (undeformed) strings would similarly naturally require quantization of the inverse string temperature $R$, arising from factors of $e^{-R E}$. These quantizations of $R$ and $J$ with respect to $E$ and $\tilde{E}$ respectively are precisely compatible with the length duality \eqref{eq:lengthduality}, given the relation \eqref{eq:Eandpduality}. In short, mirror duality in the TBA picture means that the thermodynamics of certain strings are related to the ground state energies (spectra) of others, and the corresponding quantization conditions arising from analyticity requirements are precisely compatible.

\section{Conclusions and outlook}

In this paper we have constructed the ground state TBA equations for strings on the $\left(\ads\right)_\eta$ background. Our construction relies on the assumption that
the $q$-deformed scattering matrix found by requiring quantum group symmetry is the exact S-matrix for strings on $\left(\ads\right)_\eta$ for finite values of the coupling constant.
In the process of investigating the mirror theory obtained from the original one by a double Wick rotation, we have found an interesting duality transformation.
This transformation relates a mirror theory at a fixed value of the deformation parameter $\r$ to a string theory at $\pi-\r$. This property is respected by the dispersion relations, S-matrices, and the Bethe-Yang equations. Also semi-classical considerations are compatible with the existence of this duality. This setup is further corroborated by the giant magnon solutions we have constructed for the deformed string and mirror theories. Interestingly, in the spirit of Zamolodchikov, mirror duality means that we can study the spectra of certain strings via the thermodynamic properties of others, for example raising the question whether concepts such as the Hagedorn temperature simply fit into this story or can be used to gain interesting new insights.

\smallskip

There are several directions extending our investigations. First, it would be interesting to understand how the duality transformations acts at the level of the full sigma-model, \emph{i.e.} without restriction to a single-field case. The main difficulty here lies in constructing the field redefinitions that would recast the dualized string action in the form of the mirror action. Next, the analysis of giant magnons could be extended to the finite size case to gain insight on the finite-size corrections to the dispersion relation at strong coupling. It should also be possible to apply the recently developed quantum spectral curve method \cite{Gromov:2013pga,Cavaglia:2014exa,Gromov:2014eha} to the models at hand, and obtain further insights into the finite-size spectrum. In particular, it would be fascinating to shed light on the functional form of $q(\eta, g)$ that enters the exact S-matrix of the $\eta$-deformed model. Also, as already indicated it would be exciting if the limit $\r\rightarrow \pi$ can be sensibly realized at the level of the sigma model, given that this limit is so closely related to the $\ads$ mirror model.

\smallskip

Finally, the deformation of the classical sigma model we built on belongs to a general class of deformations governed by solutions of the classical Yang-Baxter equation. Among these there are also deformations of so-called Jordanian type. Recently, these deformations have been considered in the context of the string sigma model in \cite{Kawaguchi:2014qwa,Kawaguchi:2014fca}. It would be interesting to attempt to generalize our considerations to these Jordanian deformations as well. Of course, it would be very interesting if these deformations can be translated through AdS/CFT.

\section*{Note added}

While the work presented in this paper was already finished \cite{stijnslondontalk}, the interesting paper \cite{Hoare:2014pna} appeared, discussing the limit $\varkappa \rightarrow \infty$ ($\nu\rightarrow1$) of the deformed background geometry. There it is shown that upon appropriate rescalings the deformed geometry is T-dual to that of $\mathrm{dS}_5 \times \mathrm{H}^5$, which can be obtained by a ``double Wick rotation'' from $\ads$. To avoid confusion let us emphasize that this double Wick rotation acts on the target space fields, and is hence a priori different from the Wick rotations of world-sheet coordinates that we are discussing. Of course, in the light-cone gauge the two concepts can become intertwined, but note that the $\ads$ mirror theory is unitary, while the authors of \cite{Hoare:2014pna} indicate problems with unitarity in their case. To our current knowledge the Wick rotation statements of \cite{Hoare:2014pna} refer simply to the analytic continuation of $H^5$ and $\mathrm{dS}_5$ to $\mathrm{AdS}_5$ and $S^5$ respectively. Let us also note that the possible non-unitarity indicated in \cite{Hoare:2014pna} (as discussed on their page 7), may have a natural explanation from our point of view, as discussed below eqn. \eqref{eq:unitarityregime}.  Given the interesting nature of the limit $\nu \rightarrow 1$ ($\r \rightarrow \pi$) discussed in the main text, it would clearly be interesting to further investigate the relation between the results of \cite{Hoare:2014pna} and our present work.

\section*{Acknowledgements}

We would like to thank N. Beisert, R. Borsato and S. Frolov for discussions and R. Borsato and S. Frolov for useful comments on the manuscript. ST is supported by the Einstein Foundation Berlin in the framework of the research project "Gravitation and High Energy Physics" and acknowledges further support from the People Programme (Marie Curie Actions) of the European Union's Seventh Framework Programme FP7/2007-2013/ under REA Grant Agreement No 317089. G.A. acknowledges support by the Netherlands Organization for Scientific Research (NWO) under the VICI grant 680-47-602. The work by G.A. is also a part of the ERC Advanced grant research programme No. 246974,  {\it ``Supersymmetry: a window to non-perturbative physics"} and of the D-ITP consortium, a program of the NWO that is funded by the Dutch Ministry of Education, Culture and Science (OCW). The work of ML is partially supported by grant no.\ 200021-137616 from the Swiss National Science Foundation.


\appendix

\section{Appendices}

\subsection{The S-matrix}
\label{app:qdefmatrixSmatrix}

Here we briefly present the $\mathfrak{psu}_q(2|2)_{c.e.}$ invariant S-matrix. We use $E_{ij}$ to denote the $4\times 4$ matrix with a one in entry $(i,j)$ and zeroes everywhere else. Next, we introduce the following definition
\begin{equation}
E_{kilj}=(-1)^{\epsilon(l)\epsilon(k)}E_{ki}\otimes E_{lj}\, ,
\end{equation}
where $\epsilon(i)$ denotes the parity of the index, zero for $i=1,2$ (bosons) and one for $i=3,4$ (fermions). The matrices $E_{kilj}$
can be used to write down invariants with respect to the action of two copies of $\su_{q}(2)$. If we introduce
\begin{align}
\Lambda_1=&E_{1111}+\frac{q}{2}E_{1122}+\frac{1}{2}(2-q^2)E_{1221}+\frac{1}{2}E_{2112}+\frac{q}{2}E_{2211}+E_{2222}\, ,\nonumber\\
\Lambda_2=&\frac{1}{2}E_{1122}-\frac{q}{2}E_{1221}-\frac{1}{2q}E_{2112}+\frac{1}{2}E_{2211}\, , \nonumber \\
\Lambda_3=&E_{3333}+\frac{q}{2}E_{3344}+\frac{1}{2}(2-q^2)E_{3443}+\frac{1}{2}E_{4334}+\frac{q}{2}E_{4433}+E_{4444} \, , \nonumber\\
\Lambda_4=&\frac{1}{2}E_{3344}-\frac{q}{2}E_{3443}-\frac{1}{2q}E_{4334}+\frac{1}{2}E_{4433}\, , \nonumber\\
\Lambda_5=&E_{1133}+E_{1144}+E_{2233}+E_{2244}\, ,\\
\Lambda_6=&E_{3311}+E_{3322}+E_{4411}+E_{4422}\, , \nonumber\\
\Lambda_7=&E_{1324}-qE_{1423}-\frac{1}{q}E_{2314}+E_{2413}\, , \nonumber\\
\Lambda_8=&E_{3142}-qE_{3214}-\frac{1}{q}E_{4132}+E_{4231}\, , \nonumber\\
\Lambda_9=&E_{1331}+E_{1441}+E_{2332}+E_{2442}\, , \nonumber\\
\Lambda_{10}=&E_{3113}+E_{3223}+E_{4114}+E_{4224}\, , \nonumber
\end{align}
the $\mathfrak{psu}_q(2|2)_{c.e.}$ S-matrix is given by
\begin{equation}
S_{12}(p_1,p_2)=\sum_{k=1}^{10}a_k(p_1,p_2)\Lambda_k\, ,
\end{equation}
where the coefficients are
\begin{align}
a_1=&1\, ,  \nonumber \\
a_2=&-q+\frac{2}{q}\frac{x^-_1(1-x^-_2x^+_1)(x^+_1-x^+_2)}{x^+_1(1-x^-_1x^-_2)(x^-_1-x^+_2)}\nonumber \\
a_3=&\frac{U_2V_2}{U_1V_1}\frac{x^+_1-x^-_2}{x^-_1-x^+_2} \\
a_4=&-q\frac{U_2V_2}{U_1V_1}\frac{x^+_1-x^-_2}{x^-_1-x^+_2}+\frac{2}{q}\frac{U_2V_2}{U_1V_1}\frac{x^-_2(x^+_1-x^+_2)(1-x^-_1x^+_2)}{x^+_2(x^-_1-x^+_2)(1-x^-_1x^-_2)}\nonumber \\
a_5=&\frac{x^+_1-x^+_2}{\sqrt{q}\, U_1V_1(x^-_1-x^+_2)}\nonumber
\end{align}
\begin{align}
a_6=&\frac{\sqrt{q}\, U_2V_2(x^-_1-x^-_2)}{x^-_1-x^+_2} \nonumber\\
a_7=&i\frac{(x^+_1-x^-_1)(x^+_1-x^+_2)(x^+_2-x^-_2)}{\sqrt{q}\, U_1V_1 (1-x^-_1 x^-_2) (x^-_1-x^+_2)\gamma_1\gamma_2}
\nonumber \\
a_8=&i\frac{U_2V_2\,  x^-_1x^-_2(x^+_1-x^+_2)\gamma_1\gamma_2}{q^{\frac{3}{2}} x^+_1x^+_2(x^-_1-x^+_2)(x^-_1x^-_2-1)}\\
a_9=&\frac{(x^-_1-x^+_1)\gamma_2}{(x^-_1-x^+_2)\gamma_1} \nonumber \\
\nonumber
a_{10}=&\frac{U_2V_2 (x^-_2-x^+_2)\gamma_1}{U_1V_1(x^-_1-x^+_2)\gamma_2}\, .
\end{align}
The central charges are given in eqn. \eqref{eq:Eandpdef}, and the parameters $\gamma_i$ are
\begin{equation}
\gamma_i=q^{\frac{1}{4}}\sqrt{i(x^-_i-x^+_i)U_iV_i}\, .
\end{equation}
It is important to note that the dependence of the S-matrix on the variables $\gamma_i$, $i=1,2$, is gauge-like. Indeed, introducing the diagonal matrix
$\Gamma_i={\rm diag}(1,1,\gamma_i,\gamma_i)$, we find
\begin{equation}
\Big[ \Gamma_1\otimes \Gamma_2\Big]\,  S_{12}^{\gamma_i=1}(z_1,z_2)\, \Big[\Gamma_1^{-1}\otimes \Gamma_2^{-1}\Big]=S_{12}(z_1,z_2)\, ,
\end{equation}
where $S_{12}^{\gamma_i=1}$ is the above S-matrix with $\gamma_1$ and $\gamma_2$ set to one. This means that at the level of the spectrum we can forget about the $\gamma$ factors, which is what is alluded to when discussing the invariance and inversion properties of the S-matrix in section \ref{sec:parametrization}, as these are affected by the branch choices in the $\gamma$ factors. Note that for coincident arguments the S-matrix reduces to the (graded) permutation.


\subsection*{Excitation numbers}
We also give here the relationship between the unbroken isometries of our deformed model and excitation numbers appearing in the Bethe-Yang equations
(\ref{eq:BYstring})-(\ref{eq:auxBEw}).
These excitation numbers are related to the global conserved charges as
\begin{align}
J_1 &= J\,, & J_2 &= \frac{q_l + q_r}{2}\,, & J_3 &= \frac{q_l - q_r}{2}\,,\\
H_1 &= H\,, & H_2 &= \frac{s_l + s_r}{2}\,, & H_3 &= \frac{s_l - s_r}{2}\,,
\end{align}
with
\begin{equation}
q_\alpha = K^{\mathrm{I}} - K^{\mathrm{II}}_\alpha \, , \,\, \mbox{ and }\,\, s_\alpha = K^{\mathrm{II}}_{\alpha} - 2K^{\mathrm{III}}_\alpha\, .
\end{equation}
Here the $J_i$ are conserved $U(1)$ charges that become the angular momenta on $S^5$ in the undeformed limit, and $H_2$ and $H_3$ are the analogous quantities on deformed anti-de Sitter space (the $s_\alpha$ would be spins). $H$ would be the target space energy related to the world-sheet energy $E$ as $H=E-J$.

\subsection*{Momentum and energy}

The energy and momentum of the deformed string sigma model are expressed in terms of $u$ as
\bea
p=\frac{1}{i}\log\frac{\sin\frac{1}{2}(u+ia)}{\sin\frac{1}{2}(u-ia)}\frac{1+\sqrt{\frac{\cos(u+ia)-\cos\r}{\cos(u+ia)-1}}}{1+\sqrt{\frac{\cos(u-ia)-\cos\r}{\cos(u-ia)-1}}}\, ,
\eea
and
\bea
H=\frac{1}{a}\log \frac{ \cos\tfrac{1}{2}(u+ia)-i \sin\tfrac{1}{2}(u+ia) \sqrt{\frac{\cos(u+ia)-\cos\r}{\cos(u+ia)-1}}    }
{    \cos\tfrac{1}{2}(u-ia)-i \sin\tfrac{1}{2}(u-ia) \sqrt{\frac{\cos(u-ia)-\cos\r}{\cos(u-ia)-1}}    }\, .
\eea

\subsection{The $q$-deformed dressing phase}
\label{app:dressingphase}

As introduced in eqn. \eqref{eq:Ssu2def} the dressing phase needs to satisfy the crossing equation
\begin{equation}
\label{eq:cross2}
\sigma(z_1,z_2)\sigma(z_1,z_2-\omega_2) = q^{-1}\frac{x_1^+ + \xi}{x_1^- + \xi}\frac{x_1^--x_2^+}{x_1^--x_2^-}\frac{1-\frac{1}{x_1^+x_2^+}}{1-\frac{1}{x_1^+x_2^-}}\, ,
\end{equation}
which by unitarity is equivalent to
\begin{equation}
\label{eq:cross1}
\sigma(z_1+\omega_2,z_2)\sigma(z_1,z_2)=q^{-1}\frac{x_2^- + \xi}{x_2^+ + \xi} \frac{x_1^--x_2^+}{x_1^--x_2^-}\frac{1-\frac{1}{x_1^+x_2^+}}{1-\frac{1}{x_1^+x_2^-}}\, .
\end{equation}
To obtain a natural deformation of the $\ads$ dressing phase \cite{Beisert:2006ez} we can follow the methods of \cite{Volin:2009uv}, as already done in the phase deformed context in \cite{Hoare:2011wr} (see also appendix D of \cite{qTBAII}). The undeformed dressing phase is conventionally written in the form \cite{Arutyunov:2004vx}
\begin{equation}
\label{eq:dressingphasechidef}
\sigma(z_1,z_2) \equiv e^{i \theta_{12}}= \exp { i\left(\chi(x_1^+,x_2^+) - \chi(x_1^-,x_2^+) - \chi(x_1^+,x_2^-) + \chi(x_1^-,x_2^-)\right) } \, ,
\end{equation}
where when both particles are in the string region, the $\chi$-functions are given by \cite{Dorey:2007xn}
\begin{equation}
\nonumber
\chi(x_1,x_2) = \Phi(x_1,x_2) \equiv i \oint_{|z|=1} \frac{dz}{2 \pi i} \frac{1}{z-x_1} \oint_{|w|=1}\frac{dw}{2 \pi i}  \frac{1}{w-x_2} \log  \frac{\Gamma(1+\frac{ig}{2}(u(z)-u(w)))}{\Gamma (1-\frac{ig}{2}(u(z)-u(w)))}\,.
\end{equation}
In the picture of \cite{Volin:2009uv}, the ratio of $\Gamma$-functions arises as the solution to a difference equation, while the integration contours arise as the values of the $x_s$-function along its cuts on the $u$-plane. In the $q$-deformed context, the deformed difference equation is naturally solved by a ratio of $\Gamma_{q^2}$ functions \cite{Hoare:2011wr}, and while in the phase deformed context the integration contours stay the same, here the integration contours are modified.\footnote{Of course for some (most) external arguments these deformed contours may be homotopically equivalent to unit circles. However, only the modified integration contours result in a nice analytic structure analogous to the undeformed dressing phase.}
As a result, our $\Phi$-functions are given by
\begin{equation}
\label{eq:phidef}
\Phi(x_1,x_2) = i \oint_{\mathcal{C}} \frac{dz}{2 \pi i} \frac{1}{z-x_1} \oint_{\mathcal{C}}\frac{dw}{2 \pi i}   \frac{1}{w-x_2}\log  \frac{\Gamma_{q^2}(1+\frac{i}{2a}(u(z)-u(w)))}{\Gamma_{q^2}(1-\frac{i}{2a}(u(z)-u(w)))}\,.
\end{equation}
where we have rescaled the rapidities in accordance with our conventions, and $z$ and $w$ run along the contour specified by the solution to the equation
\begin{equation}
\label{eq:dressingcontourdef}
\mathcal{C} \, : \,\,|y|^2-1+(y^*-y)\xi = 0\,.
\end{equation}
We can parametrize this circle of radius $\sqrt{1-\xi^2}$ with its center shifted by $\xi$ as such, giving
\begin{equation}
y=\sqrt{1-\xi^2}e^{i\varphi} + \xi\, ,
\end{equation}
or we can substitute polar coordinates (for $-y$) to get
\begin{equation}
y=\left(\sqrt{1-\sin^2 \tfrac{\r}{2} \cos^2\phi }- \sin \tfrac{\r}{2} \sin \phi\right)\frac{e^{i\phi}}{\cos\frac{\r}{2}}\, .
\end{equation}
The rapity along this contour is then given by
\begin{align}
u(y(\varphi)) & = i \log \frac{1+i \sin \frac{\r}{2} \, e^{i \varphi}}{1-i \sin \frac{\r}{2} \, e^{-i \varphi}}\, ,\\
u(y(\phi)) & = -2 \arcsin (\sin \tfrac{\r}{2} \cos \phi)\, , \label{eq:rapiditycontourparam2}
\end{align}
as follows by its definition \eqref{eq:rapiditydef}. Both parametrizations reduce directly to the standard parametrization of the circle and, upon rescaling, the rapidity in the limit $\r\rightarrow0$.

Let us note that the numerical mismatch reported in \cite{Arutyunov:2013ega} between the semi-classical expansion of this dressing phase and the explicit sigma model result for $\nu>1/\sqrt{2}$
appears to be due to an unfortunate branch choice obtained by combining the differences of rapidities entering in \eqref{eq:phidef} under a single logarithm via their definition \eqref{eq:rapiditydef}. While this is harder to observe when using unit circles as integration contours, in terms of the deformed contours we can readily see that the single logarithm picks up a cut for $\nu>1/\sqrt{2}$. With our current conventions we perfectly match the scattering phase of the semi-classical sigma model.

\subsubsection*{Properties of the dressing phase}

The dressing phase has two simple properties that readily follow from the integral representation of the $\Phi$-function in the `string-string' region, and elsewhere by analytic continuation. Firstly the $\Phi$-function is invariant under a change of sign of $\xi$
\begin{equation}
\Phi(x_1,x_2,-\xi,q)=\Phi(x_1,x_2,\xi,q)\,.
\end{equation}
To show this, we simply note that both the $x$-functions and the integration contour flip sign when $\xi$ does. This means that after changing variables in $\Phi(x_1,x_2,-\xi,q)$ from $z$ and $w$ to $\tilde{z}=-z$ and $\tilde{w}=-w$ respectively, and identifying $x_1$, $x_2$ and $u$ (cf. eqn. \eqref{eq:rapiditydef}) in terms of their left hand side counterparts, we get exactly $\Phi(x_1,x_2,\xi,q)$. From this it immediately follows that the entire dressing phase is invariant under a change of sign on $\xi$. Secondly, the $\Phi$-function picks up a sign when inverting $q$
\begin{equation}
\Phi(x_1,x_2,\xi,1/q)=-\Phi(x_1,x_2,\xi,q)\,.
\end{equation}
This follows by relating $\Gamma_{q^{-2}}$ to $\Gamma_{q^2}$ via
\begin{equation}
\Gamma_{q^{-2}}(z)=q^{-z^2+3z-2}\Gamma_{q^2}(z)\,,
\end{equation}
and noting that apart from this the arguments of the two $\Gamma_{q^2}$ functions effectively interchange, so that we get
\begin{align}
& \Phi(x_1,x_2,\xi,1/q) = \\
& \hspace{20pt} i \oint \frac{dz}{2 \pi i} \frac{1}{z-x_1} \oint \frac{dw}{2 \pi i}   \frac{1}{w-x_2}\left(\log  \frac{\Gamma_{q^2}(1-\frac{i}{2a}(u(z)-u(w)))}{\Gamma_{q^2}(1+\frac{i}{2a}(u(z)-u(w)))} - i(u(z)-u(w)) \right)\,, \nonumber
\end{align}
where as always $a = \log q$. The first part of this integral is nothing but $-\Phi(x_1,x_2,\xi,q)$, while the extra $u(z)-u(w)$ term integrates to zero since in the contour in the $w$ integration of $u(z)$ can be shrunk to nothing, and vice versa. Of course in the dressing phase the arguments of the $\Phi$-functions depend on $q$, but since $x^+$ and $x^-$ interchange under inversion of $q$ this does not affect the combination entering the dressing phase (cf. eqn. \eqref{eq:dressingphasechidef}), meaning it inverts under inversion of $q$. Another property of the dressing phase is that it inverts under a parity transformation, \emph{i.e.}
\begin{equation}
\label{eq:dressingparity}
\sigma(-u,-v;h,a) = \sigma^{-1}(u,v;h,a)\, .
\end{equation}
Due to the more involved parity transformation properties of the $q$-deformed $x$-functions this is not an obvious statement. Still it is easy to see that the crossing equation inverts under a parity transformation, with the crossing transformation then implemented in the opposite direction on the torus of course. Eqn. \eqref{eq:dressingparity} can of course be readily verified numerically.

\subsubsection*{The mirror dressing phase}

To obtain the mirror dressing phase from the above expressions, we need to analytically continue the $\chi$ functions out of the string region, or in other words through the cuts of the $x$-functions. At the level of the $x$-variables the relevant distinction is whether they lie in the interior or exterior of $\mathcal{C}$. Let us label these regions $\mathcal{R}$ by a lower index for each particle, where the lower index takes value $0$ if both the associated $x^-$ and $x^+$ variable lie in the interior and value $1$ if $x^-$ is in the interior but $x^+$ is in the exterior. The line of real momenta in the mirror theory is contained in the region $\mathcal{R}_{1,1}$ and will be of primary interest to us, we will not need the remaining cases here.

In terms of the above integral representation, if we move from $\mathcal{R}_{0,0}$ to $\mathcal{R}_{1,0}$ or $\mathcal{R}_{0,1}$ the pole of the integrand at $z=x_1$ respectively $w=x_2$ crosses the integration contour. Concretely, the discontinuity of $\Phi$ in its first argument is given by
\begin{equation}
\label{eq:psidef}
\Psi(x_1,x_2) \equiv i \oint_{\mathcal{C}} \frac{dz}{2 \pi i} \frac{1}{z-x_2} \log  \frac{\Gamma_{q^2} (1+\frac{i}{2a}(u_1-u(z)))}{\Gamma_{q^2} (1-\frac{i}{2a}(u_1-u(z)))} \, .
\end{equation}
The discontinuity in the second argument is analogous. This $\Psi$-function has discontinuities in both its first and second arguments, and in particular to get to $\mathcal{R}_{1,1}$ from $\mathcal{R}_{1,0}$ we need to add
\begin{equation}
i\log  \frac{\Gamma_{q^2} (1+\frac{i}{2a}(u_1-u_2))}{\Gamma_{q^2} (1-\frac{i}{2a}(u_1-u_2))} \, .
\end{equation}
In total the analytic continuation to $\mathcal{R}_{1,1}$ results in the $\chi$ functions
\begin{align}
\nonumber \chi(x_1^+, x_2^+) =&\,
\Phi(x_1^+, x_2^+)+ \Psi(x_2^+, x_1^+) - \Psi(x_1^+,
x_2^+)\\\nonumber &\hspace{20pt} +i \log\frac{\Gamma_{q^2} (1+\frac{ig}{2}(x_1^++\frac{1}{x_1^+}-x_2^+-\frac{1}{x_2^+}))}
{\Gamma_{q^2} (1-\frac{ig}{2}(x_1^++\frac{1}{x_1^+}-x_2^+-\frac{1}{x_2^+}))}\,,\\\nonumber
\chi(x_1^+, x_2^-) =&\, \Phi(x_1^+, x_2^-) - \Psi(x_1^+, x_2^-)
\,,\\
\chi(x_1^-, x_2^+) =&\, \Phi(x_1^-, x_2^+)+ \Psi(x_2^+, x_1^-)\,,\nonumber\\
\chi(x_1^-, x_2^-) =&\, \Phi(x_1^-, x_2^-)\,.
\end{align}
We should note that the resulting mirror dressing `phase' is not unitary, while of course the full mirror scalar factor and in particular also the improved mirror dressing phase are. Using these expressions we can easily check mirror duality of the dressing phase in the form of eqn. \eqref{eq:dressingduality} numerically.

\subsection{Gauge-fixed action}
\label{app:GFA}
In this appendix we fix the uniform light-cone gauge and find the corresponding gauge-fixed action. For the case of $\ads$ this has been done in \cite{Klose:2006zd}, here we will use a different method.

We start by considering a generic action of the form
\begin{equation}\label{action}
S=-\frac{g}{2} \int_{-r}^r \, {\rm d}\sigma {\rm d} \tau \left( \, \gamma^{\alpha\beta} \partial_\alpha X^M \partial_\beta X^N G_{MN} -\epsilon^{\alpha\beta} \partial_\alpha X^M \partial_\beta X^N B_{MN} \right),
\end{equation}
where $G_{MN}$ and $B_{MN}$ are the background metric and $B$-field respectively.
This action can be rewritten in terms of light-cone coordinates as
\begin{equation}
S=  \int_{-r}^r \, {\rm d}\sigma {\rm d} \tau \left( p_+ \dot{x}^+ +p_- \dot{x}^- +p_{\mu}\dot{x}^{\mu} + \frac{\gamma^{01}}{\gamma^{00}} C_1 + \frac{1}{2g \gamma^{00}} C_2  \right).
\end{equation}
Here $C_1$ and $C_2$ are
\bea
C_1 &=&p_- x'^{-}+p_+x'^{+}+p_{\mu}x'^{\mu}, \\
C_2 &=& G^{--} p_-^2 +2 G^{+-} p_+ p_- + G^{++} p_+^2 \\
&&~+ g^2 G_{--} x'^- x'^- + 2 g^2 G_{+-} x'^+x'^- + g^2 G_{++} x'^+x'^+ + \mathcal{H}_x\, ,\nonumber
\eea
where we have taken into account that for the background we consider $B^{+-}=0$. Diffeomorphism invariance results in the Virasoro constraints $C_i=0$. The light-cone components of the inverse metric are
\begin{equation*}
G^{++}=\frac{G_{--}}{G_{++}G_{--}-G_{+-}^2}\, , ~~~G^{+-}=-\frac{G_{+-}}{G_{++}G_{--}-G_{+-}^2}\, , ~~~
G^{--}=\frac{G_{++}}{G_{++}G_{--}-G_{+-}^2}\, ,
\end{equation*}
and the quantity $\mathcal{H}_x$ is the part that depends on the transverse fields only
\begin{equation}
\mathcal{H}_x = G^{\mu\nu} p_\mu p_\nu + g^2 x'^\mu x'^\nu G_{\mu\nu} -2 g p_\mu x'^{\rho} G^{\mu\nu} B_{\nu\rho} + g^2 x'^\lambda x'^\rho B_{\mu\lambda} B_{\nu\rho} G^{\mu\nu}.
\end{equation}
Next we impose the light-cone gauge
\begin{equation}
x^+= \tau, \qquad p_-=1.
\end{equation}
Then the Hamiltonian density ${\cal H}=-p_+$ follows from the constraint $C_2=0$
\bea
\hspace{-0.5cm} \mathcal{H}=\frac{1}{G^{++}}\Bigg[G^{+-} +\sqrt{(G^{+-2}-G^{++}G^{--})-G^{++}( g^2 G_{--} x'^- x'^- + \mathcal{H}_x)}\Bigg]\, .
\eea
We write this Hamiltonian in the form
\bea
\mathcal{H}=\frac{G^{+-}}{G^{++}} +\frac{1}{G^{++}}\sqrt{\mathcal{W}}\, ,\eea
where
\bea
\nonumber
\mathcal{W}&=&(G^{+-2}-G^{++}G^{--})-G^{++}\Big[ ( G^{\mu\nu} +g^2 G_{--} x'^{\mu}x'^{\nu})p_{\mu}p_{\nu}+\nonumber \\
\nonumber
&+& g^2 x'^\mu x'^\nu G_{\mu\nu} -2 g p_\mu x'^{\rho} G^{\mu\nu} B_{\nu\rho} + g^2 x'^\lambda x'^\rho B_{\mu\lambda} B_{\nu\rho} G^{\mu\nu}\Big]\, .\eea
Next, the Hamiltonian equations of motion for $x^{\mu}$ read
\bea\nonumber
\dot{x}^{\mu}=\frac{\pa \mathcal{H}}{\pa p_{\mu}}=-\frac{1}{\sqrt{\mathcal W}}\Bigg[( G^{\mu\nu} +g^2 G_{--} x'^{\mu}x'^{\nu})p_{\nu} - g x'^{\rho} G^{\mu\nu} B_{\nu\rho}\Bigg]\, .
\eea
We can write them in the form
\bea
\label{cm}
M^{\mu\nu}p_{\nu}=-\sqrt{\mathcal W} ~\dot{x}^{\mu}+g G^{\mu\tau} x'^{\sigma}  B_{\tau\sigma}\, ,
\eea
where we have introduced the matrix $M$ with entries
$$
M^{\mu\nu}=G^{\mu\nu} +g^2 G_{--} x'^{\mu}x'^{\nu}\, .
$$
From (\ref{cm}) we find the canonical momenta in terms of velocities
\bea
p_{\mu}=-M^{-1}_{\mu\nu}( \sqrt{\mathcal W} ~\dot{x}^{\nu} - g G^{\nu\tau} x'^{\sigma}  B_{\tau\sigma})\, .
\eea
Upon substituting this expression for the momentum back into $\mathcal{W}$, we obtain an equation for $\mathcal{W}$
\bea\nonumber
\mathcal{W}&=&(G^{+-2}-G^{++}G^{--})-G^{++}\Bigg[ \mathcal{W}~M^{-1}_{\mu\nu}\dot{x}^{\mu}\dot{x}^{\nu}-\\
&&-g^2 M^{-1}_{\mu\nu}G^{\mu\tau} G^{\nu\sigma}x'^{\alpha}x'^{\beta}B_{\tau\alpha}B_{\sigma\beta}+g^2 x'^\mu x'^\nu G_{\mu\nu}
+ g^2 x'^\lambda x'^\rho B_{\mu\lambda} B_{\nu\rho} G^{\mu\nu}\Bigg]\, ,
\nonumber\eea
which has the solution
\bea\nonumber
\mathcal{W}&=&\frac{1}{1+G^{++}M^{-1}_{\mu\nu}\dot{x}^{\mu}\dot{x}^{\nu}}\Bigg[
(G^{+-2}-G^{++}G^{--})-\\
&&-G^{++} \Big(-g^2 M^{-1}_{\mu\nu}G^{\mu\tau} G^{\nu\sigma}x'^{\alpha}x'^{\beta}B_{\tau\alpha}B_{\sigma\beta}
+g^2 x'^\mu x'^\nu G_{\mu\nu} + g^2 x'^\lambda x'^\rho B_{\mu\lambda} B_{\nu\rho} G^{\mu\nu}\Big)
\Bigg]\, .  \nonumber
\eea
Hence, the gauge-fixed Lagrangian is
\bea\nonumber
\mathscr{L}=p_{\mu}\dot{x}^{\mu}-\mathcal{H}=-\frac{G^{+-}}{G^{++}}+gM^{-1}_{\mu\nu}G^{\nu\tau}\dot{x}^{\mu}x'^{\sigma}B_{\tau\sigma}-\frac{1}{G^{++}}(1+G^{++}M^{-1}_{\mu\nu}\dot{x}^{\mu}\dot{x}^{\nu})\sqrt{\mathcal{W}}\, .
\eea
Finally, the matrix $M$ can be inverted by using the Sherman-Morrison formula
\begin{equation*}
M^{-1}_{\mu\nu}=G_{\mu\nu}-g^2G_{--}\frac{G_{\mu\tau}G_{\nu\sigma} x'^{\tau}x'^{\sigma}}{1+g^2 G_{--} G_{\tau\sigma}x'^{\tau}x'^{\sigma} }\, .
\end{equation*}
We can use this formula to bring the Lagrangian to the form
\bea\nonumber
\mathscr{L}=-\frac{G^{+-}}{G^{++}}-\frac{1}{G^{++}}\Big(1+G^{++}M^{-1}_{\mu\nu}\dot{x}^{\mu}\dot{x}^{\nu}\Big)\sqrt{\mathcal{W}}+g\dot{x}^{\mu}x'^{\nu}B_{\mu\nu}\, ,
\eea
where
\bea\nonumber
\mathcal{W}=(G^{+-2}-G^{++}G^{--}) \frac{1+g^2 G_{--}  G_{\mu\nu}  x'^\mu x'^\nu  }{1+G^{++}M^{-1}_{\mu\nu}\dot{x}^{\mu}\dot{x}^{\nu}}\, .
\eea
To treat the spatial and temporal derivatives on equal footing we perform a rescaling $\sigma\to g \sigma$ so that upon final simplification the gauged-fixed action takes the form
\bea
\label{gfaction}
S=g\int\limits_{-r/g}^{r/g}{\rm d}\tau{\rm d}\sigma \, \mathscr{L}\, ,
\eea
where
\bea
\label{gfLag}
\mathscr{L}=-\frac{G^{+-}}{G^{++}}  -\sqrt{-\frac{\mathscr X}{G^{++}G_{--}}  }  + \dot{x}^{\mu}x'^{\nu}B_{\mu\nu}\,
\eea
and
\bea\nonumber
{\mathscr X}=1+G^{++}G_{\mu\nu}\dot{x}^{\mu}\dot{x}^{\nu}&+&G_{--}G_{\mu\nu}x'^{\mu}x'^{\nu}+\\
\label{inX}
&+&G^{++}G_{--}(G_{\mu\nu}G_{\tau\sigma}-G_{\mu\tau}G_{\nu\sigma})
\dot{x}^{\mu}\dot{x}^{\nu}x'^{\tau}x'^{\sigma}\, .
\eea
This is a gauge-fixed action for an arbitrary background $(G_{MN}, B_{MN})$ with vanishing light-cone components of $B$.  Note that in the limit $r\to\infty$
the coupling constant $g$ enters into the action just as an overall scaling factor. Furthermore, the part of the Lagrangian which contains the $B$-field, as well as the last term in eqn. \eqref{inX} are invariant under the double Wick rotation. Finally, it is easy to see that taking a flat Minkowski metric the action reduces to the standard one for eight free bosons.

\medskip

Now we are ready to specify the general form of the gauge-fixed action for the string on $\eta$-deformed ${\rm AdS}_5\times {\rm S}^5$.
Before gauge fixing the light-cone Lagrangian corresponding to (\ref{action}) is
\bea
\L&=&-\frac{g}{2}\gamma^{\a\b}\Big[G_{++}\pa_{\a}x^+ \pa_{\b}x^+ + 2G_{+-}\pa_{\a}x^+ \pa_{\b}x^- + G_{--}\pa_{\a}x^- \pa_{\b}x^-+\nonumber\\
&&~~~~~~~+ G_{ij}^{\alg{a}}\pa_{\a}z^i \pa_{\b}z^j+G_{ij}^{\alg{s}}\pa_{\a}y^i \pa_{\b}y^j\Big]+\mathscr{L}^{WZ}\, .\eea
Here $x^{\mu}\in \{z^i,y^i\}$, where $z^i$ and $y^i$ with $i=1,\ldots, 4$ are the transverse coordinates of the deformed AdS and five-sphere respectively, while $\mathscr{L}^{WZ}$ is the Wess-Zumino term comprising the contribution of the $B$-field. The light-cone coordinates $x^{\pm}$ are defined as
\bea
x^+=(1-\agauge)t+\agauge\phi\, , ~~~~x^{-}=\phi-t\, ,
\eea
where $\agauge$ is a parameter of the generalized light-cone gauge.
The light-cone components of the metric are
\bea\nonumber
G_{++}&=&G_{\phi\phi}-G_{tt} \, , ~~~
G_{+-}=\agauge G_{tt}+(1-\agauge) G_{\phi\phi}\, , \\
G_{--}&=&(1-\agauge)^2 G_{\phi\phi}-\agauge^2 G_{tt}\, ,\eea
and
\bea\nonumber
G^{++}&=&\agauge^2G_{\phi\phi}^{-1}+(1-\agauge)^2G_{tt}^{-1} \, , ~~~
G^{+-}=\agauge G_{\phi\phi}^{-1}+(1-\agauge) G_{tt}^{-1}\, , \\
G^{--}&=&G_{\phi\phi}^{-1}-\agauge^2 G_{tt}^{-1}\, .
\eea
Below we have collected the other components of the metric including $G_{tt}$ and $G_{\phi\phi}$ as well as the Wess-Zumino term.

\subsubsection*{Metric and Wess-Zumino term}

Let the coordinates $z_i$, $i=1,\ldots,4$, and $t$ parametrize the deformed AdS space, while the coordinates $y_i$, $i=1,\ldots,4$, and the angle $\phi$ parametrize the deformed five-sphere.
Introduce the following auxiliary functions for the AdS part
\begin{align}
\nonumber
\hspace{-0.5cm}
G_{tt}&=\frac{(1+\varkappa^2)^{1/2}(1+z^2/4)^2}{(1-z^2/4)^2-\varkappa^2 z^2}\, , &
G_{zz}&=\frac{(1+\varkappa^2)^{1/2}(1-z^2/4)^2}{(1-z^2/4)^4+\varkappa^2 z^2(z_3^2+z_4^2)} \, ,
 \\
G_{\alg{a}}^{(1)}&=\frac{\varkappa^2G_{tt}G_{zz}}{(1+\varkappa^2)^{1/2}}\frac{z_3^2+z_4^2+(1-z^2/4)^2}{(1-z^2/4)^2(1+z^2/4)^2}\, , &
G_{\alg{a}}^{(2)}&=\frac{\varkappa^2 G_{zz} z^2}{(1-z^2/4)^4}
\, .
\end{align}
For the sphere part the corresponding expressions read
\begin{align}
\noindent\hspace{-0.5cm}
G_{\phi\phi}&=\frac{(1+\varkappa^2)^{1/2}(1-y^2/4)^2}{(1+y^2/4)^2+\varkappa^2 y^2}\, , &
G_{yy}&=\frac{(1+\varkappa^2)^{1/2}(1+y^2/4)^2}{(1+y^2/4)^4+\varkappa^2 y^2(y_3^2+y_4^2)} \, , \nonumber
 \\
G_{\alg{s}}^{(1)}&=\frac{\varkappa^2G_{\phi\phi}G_{yy}}{(1+\varkappa^2)^{1/2}}\frac{y_3^2+y_4^2-(1+y^2/4)^2}{(1-y^2/4)^2(1+y^2/4)^2}\, , &
G_{\alg{s}}^{(2)}&=\frac{\varkappa^2 G_{yy} y^2}{(1+y^2/4)^4}
\, .
\end{align}
In the formulae above $z^2\equiv z^iz^i$ and $y^2=y^iy^i$. In terms of these auxiliary functions, the light-cone components of the deformed AdS metric read
\begin{center}
\begin{tabular}{lll}
$G_{11}^{\alg{a}}=G_{zz}+G_{\alg{a}}^{(1)} z_1^2$,  & ~~~&~~~  $G_{33}^{\alg{a}}=G_{zz}+G_{\alg{a}}^{(1)} z_3^2+ G_{\alg{a}}^{(2)} z_4^2$,\\
$G_{22}^{\alg{a}}=G_{zz}+G_{\alg{a}}^{(1)} z_2^2$,   &  ~~~&~~~ $G_{44}^{\alg{a}}=G_{zz}+G_{\alg{a}}^{(1)} z_4^2+ G_{\alg{a}}^{(2)} z_3^2$, \\
$G_{12}^{\alg{a}}=G_{21}^{\alg{a}}=z_1 z_2 G_{\alg{a}}^{(1)}$,  & ~~~&~~~   $G_{34}^{\alg{a}}=G_{43}^{\alg{a}}=z_3z_4(G_{\alg{a}}^{(1)} -G_{\alg{a}}^{(2)} )$,
\end{tabular}
\end{center}
and analogously for the deformed sphere

\begin{center}
\begin{tabular}{lll}
$G_{11}^{\alg{s}}=G_{yy}+G_{\alg{s}}^{(1)} y_1^2$,  & ~~~&~~~  $G_{33}^{\alg{s}}=G_{yy}+G_{\alg{s}}^{(1)} y_3^2+ G_{\alg{s}}^{(2)} y_4^2$,\\
$G_{22}^{\alg{s}}=G_{yy}+G_{\alg{s}}^{(1)} y_2^2$,   &  ~~~&~~~ $G_{44}^{\alg{s}}=G_{yy}+G_{\alg{s}}^{(1)} y_4^2+ G_{\alg{s}}^{(2)} y_3^2$, \\
$G_{12}^{\alg{s}}=G_{21}^{\alg{s}}=y_1 y_2 G_{\alg{s}}^{(1)}$,  & ~~~&~~~   $G_{34}^{\alg{s}}=G_{43}^{\alg{s}}=y_3y_4(G_{\alg{s}}^{(1)} -G_{\alg{s}}^{(2)} )$.
\end{tabular}
\end{center}

\noindent
Finally, the Wess-Zumino term is given by $\mathscr{L}^{WZ}=\mathscr{L}_{\alg{a}}^{WZ}+\mathscr{L}_{\alg{s}}^{WZ}$, where
\bea
\begin{aligned}
\label{WZ}
\mathscr{L}_{\alg{a}}^{WZ}&=2g\varkappa(1+\varkappa^2)^{1\ov2}\, \eps^{\a\b}\frac{(z_3^2+z_4^2)\pa_{\a}z_1\pa_{\b}z_2}{(1-z^2/4)^4+\varkappa^2 z^2(z_3^2+z_4^2)}\, ,\\
\mathscr{L}_{\alg{s}}^{WZ}&=-2g\varkappa(1+\varkappa^2)^{1\ov2}\, \eps^{\a\b}\frac{(y_3^2+y_4^2)\pa_{\a}y_1\pa_{\b}y_2}{(1+y^2/4)^4+\varkappa^2 y^2(y_3^2+y_4^2)}\, .
\end{aligned}
\eea

\subsection{S-matrices and kernels}

\label{app:Smatricesandkernels}

Let us briefly list the energy, momentum and scattering phases the kernels entering our TBA equations are based on. The energy and momentum of mirror bound states of $Q$ particles are given by
\begin{equation}
\tilde{E}_Q(u) = -\log \frac{1}{q} \frac{x^+ + \xi}{x^- + \xi}  \, , \hspace{20pt}
\tilde{p}^Q(u) = -i a \log q \frac{x^+}{x^-} \frac{x^- +\xi}{x^+ + \xi} ,
\end{equation}
where $x^\pm$ are the mirror bound state $x$-functions $x_m(u\pm iQa)$. This follows by fusing $\tilde{U}$ and $\tilde{V}$, cf. eqn. \eqref{eq:mirrorEandpfromUandV}. Next, the basic S-matrix $S_1$ and more generally its fused forms $S_M$ and $S_{MN}$ are given by
\begin{align}
S_M(u-v) = &\, \frac{\sin{\frac{1}{2}(u - v - i M a)}}{\sin{\frac{1}{2}(u - v + i M a)}} \, ,\\
S_{MN} (u-v) = &\, S_{M+N}(u-v)S_{|M-N|}(u-v)\prod_{m=1}^{\min{(M,N)}-1}S_{|M-N|+2m}^2(u-v)\, .
\end{align}
Fusing the scattering matrix of $y^\pm$ particles with fundamental particles over a $Q$-particle bound state directly gives
\begin{align}
S_-^{yQ}(u,v)&= q^{Q/2} \, \frac{x(u) -x^-(v)}{x(u)-x^+(v)}\sqrt{\frac{x^+(v)}{x^-(v)}} \, , \\
S_+^{yQ}(u,v) &= q^{Q/2} \, \frac{\frac{1}{x(u)} -x^-(v)}{\frac{1}{x(u)}-x^+(v)}\sqrt{\frac{x^+(v)}{x^-(v)}} \, ,
\end{align}
where $x^\pm$ are the parameters for a $Q$-particle mirror bound state; $x^\pm(v) = x_m(v \pm i Q a)$. Analogously we define the S-matrices for scattering of mirror bound states with $y$-particles as
\begin{align}
S_-^{Qy}(u,v)&= q^{Q/2} \, \frac{x^-(u)-x(v)}{x^+(u)-x(v)}\sqrt{\frac{x^+(u)}{x^-(u)}} \, , \\
S_+^{Qy}(u,v) &= q^{Q/2} \, \frac{x^-(u)-\frac{1}{x(v)}}{x^+(u)-\frac{1}{x(v)}}\sqrt{\frac{x^+(u)}{x^-(u)}} \, .
\end{align}
Fusing this S-matrix over the constituents $y$-roots of an $M|vw$ string gives \cite{qTBAI}
\begin{equation}
S^{QM}_{xv}(u,v) \equiv q^{Q}\frac{x^-(u)-x^+(v) }{x^+(u)-x^+(v) }\frac{x^-(u)-x^-(v) }{x^+(u)-x^-(v)}\frac{x^+(u)}{x^-(u)}
\prod_{i=1}^{M-1} S_{Q+M-2i}(u-v) \, ,
\end{equation}
where $x^\pm(v) = x(v \pm i M a)$ and $x^\pm(u) = x(u \pm i Q a)$. The fused version of the main scalar factor, $S_{\mathfrak{sl}(2)}^{QP}$, is more involved but completely analogous to the phase-deformed case, and can be obtained from the expressions in \cite{qTBAII} (see also \cite{StijnsThesis}) by rescaling the rapidity and replacing $|q|=1$ by $q\in\mathbb{R}$ where appropriate.

\subsubsection*{Kernels}

Now let us define our integration kernels, beginning with
\begin{align}
\label{eq:KM}
K_M(u) \equiv & \, \frac{1}{2\pi i} \frac{d}{du} \log{S_M(u)} = \frac{1}{2 \pi} \frac{\sinh{M a}}{\cosh{M a}-\cos{u}}\, ,\\
K_{MN}(u) \equiv & \,\frac{1}{2\pi i} \frac{d}{du} \log{S_{MN} (u)} =K_{M+N}+K_{|M-N|} + 2 \sum_{j=1}^{\min{(M,N)}-1}K_{|M-N|+2j}\, .
\label{eq:KMN}
\end{align}
These kernels have the following properties
\begin{align}
& K_N (\delta_{N,M} - I_{NM} \star \s) =  \s\,  \delta_{M,1} \, , \label{eq:simpKM} \\
& K_{ML}(\delta_{L,N} - I_{LN} \star \s) = \s\, I_{M,N} \, ,\label{eq:simpKMN}
\end{align}
where $\s$ is defined in the main text in eqn. \eqref{eq:sigmakerneldef}. These properties readily follow by Fourier series. Next, we also have
\begin{align}
K^{QM}_{xv}(u,v) & \equiv \frac{1}{2\pi i} \frac{d}{du} \log S^{QM}_{xv} (u,v)\, ,\\
K^{MQ}_{vwx}(u,v) & \equiv - \frac{1}{2\pi i} \frac{d}{du} \log S^{QM}_{xv}(v,u)\, ,\\
K^{Qy}_{\beta}(u,v) & \equiv \frac{1}{2\pi i} \frac{d}{du} \log S^{Qy}_{\beta} (u,v) \, ,\\
K^{yQ}_{\beta}(u,v) & \equiv \beta \frac{1}{2\pi i} \frac{d}{du} \log S^{Qy}_{\beta} (v,u) \, .
\end{align}
It is convenient to consider linear combinations of the last two kernels, namely
\begin{align}
K^{Qy}_{-}(u,v) - K^{Qy}_{+}(u,v) & \equiv K_{Qy}(u,v) = K(u+iQ/g,v) - K(u-iQ/g,v)\, , \\
K^{Qy}_{-}(u,v) + K^{Qy}_{+}(u,v) & = K^{yQ}_{-}(u,v) - K^{yQ}_{+}(u,v) = K_Q(u,v)\, , \\
K^{yQ}_{-}(u,v) + K^{yQ}_{+}(u,v) & \equiv K_{yQ}(u,v) = K(u,v+iQ/g) - K(u,v-iQ/g)\, .
\end{align}
where
\begin{equation}
K(u,v) = \frac{1}{2\pi i} \frac{d}{du} \log \frac{x(u)-\frac{1}{x(v)}}{x(u)-x(v)} \, ,
\end{equation}
and $K_Q$ is defined in \eqref{eq:KM}. We can readily verify that the more involved kernels have properties similar to the basic ones above, namely
\begin{align}
K^{QN}_{xv}(\delta_{N,M}-I_{NM}\star  \s) & = \delta_{Q-1,M} \, \s + \delta_{M,1} K_{Qy}\, \hat{\star} \,  \s  \, , \label{eq:simpKQMxv} \\
K^{MP}_{vwx}(\delta_{P,Q}-I_{PQ}\star  \s) & = \delta_{M+1,Q} \, \s + \delta_{Q,1} \check{K}_M \star \s \, ,\label{eq:simpKMQvwx} \\
K_{yP}(\delta_{P,Q}-I_{PQ}\star  \s)& = \delta_{Q,1} (2\bar{K} \star \s + \, \s)\, ,\label{eq:simpKyQ}  \\
K_{P}(\delta_{P,Q}-I_{PQ}\star  \s)& = \delta_{Q,1} \, \s\, ,\label{eq:simpKQ}
\end{align}
The kernel $\bar{K}$ appearing in the above is defined as
\begin{equation}
\bar{K}(u,v) = \theta(|u|-r) \frac{1}{2\pi i} \frac{d}{du} \log \frac{x(u)-\frac{1}{x_s(v)}}{x(u)-x_s(v)} \, ,
\end{equation}
and
\begin{equation}
\check{K}_M(u,v) \equiv \bar{K}(u+i M a,v)+ \bar{K}(u-i M a,v)\, .
\end{equation}
Details on the fused dressing kernel $K_{\mathfrak{sl}(2)}^{QP}$ and the associated $\check{K}^\Sigma_Q$ can be readily obtained from the phase-deformed case \cite{qTBAI,qTBAII,StijnsThesis} upon appropriate replacements.

While we chose not to present them in this paper, the canonical TBA equations can be directly written down using these kernels, following the discussion in section \ref{sec:TBA}.

\end{document}